\documentclass[a4paper,11pt]{article}

\usepackage{amsmath,amssymb,amsthm}
\usepackage{dsfont}
\usepackage{graphicx}
\usepackage[british]{babel}
\usepackage[utf8]{inputenc}
\usepackage{xcolor}
\usepackage{authblk}
\usepackage{epstopdf}
\usepackage{cite}
\usepackage{pict2e}
\usepackage{subfigure}
\usepackage[colorlinks=true,linkcolor=blue,citecolor=blue]{hyperref}

\newtheorem{remark}{Remark}
\def\a{\lambda}
\newcommand{\R}{\mathbb{R}}
\newcommand{\be}{\begin{equation}}
\newcommand{\ee}{\end{equation}}
\newcommand{\fer}[1]{(\ref{#1})}
\newtheorem{thm}{Theorem}

\newcommand{\ff}{\hat f}
\newcommand{\fg}{\hat g}
\newcommand{\fR}{\hat f_R}
\newcommand{\fI}{\hat f_I}
\newcommand{\fS}{\hat f_S}
\newcommand{\fJ}{\hat f_J}
\newcommand{\fH}{\hat f_H}

\newcommand{\gJ}{\hat g_{J}}
\newcommand{\gH}{\hat g_{H}}
\newcommand{\hR}{h_{R}}
\newcommand{\hI}{h_{I}}
\newcommand{\hS}{h_{S}}
\newcommand{\hJ}{h_{J}}

\begin{document}
\title{Wealth distribution under the spread of infectious diseases}

%\author[1]{Author A\thanks{Corresponding Author: A.A@university.edu}}
%\author[1]{Author B\thanks{B.B@university.edu}}
%\author[1]{Author C\thanks{C.C@university.edu}}
%\author[2]{Author D\thanks{D.D@university.edu}}
%\author[2]{Author E\thanks{E.E@university.edu}}
%\affil[1]{Department of Computer Science, \LaTeX\ University}
%\affil[2]{Department of Mechanical Engineering, \LaTeX\ University}

\author[1]{G. Dimarco\thanks{\tt giacomo.dimarco@unife.it}}
 \author[1]{L. Pareschi\thanks{\tt lorenzo.pareschi@unife.it}} 
 \author[2]{G.Toscani\thanks{\tt giuseppe.toscani@unipv.it}}
 \author[2]{M.Zanella\thanks{\tt mattia.zanella@unipv.it}}

\affil[1]{\normalsize
		Mathematics and Computer Science Department,
		University of Ferrara, Italy.}
\affil[2]{
		Mathematics Department, 
		University of Pavia, Italy.}
\date{}

\maketitle

\abstract{We develop a mathematical framework to study the economic impact of infectious diseases by integrating epidemiological dynamics with a kinetic model of wealth exchange.  The multi-agent description leads to study the evolution over time of a system of kinetic equations for the wealth densities of susceptible, infectious and recovered individuals, whose proportions are driven by a classical compartmental model in epidemiology. Explicit calculations show that the spread of the disease seriously affects the distribution of wealth, which, unlike the situation in the absence of epidemics, can converge towards a stationary state with a bimodal form. Furthermore, simulations confirm the ability of the model to describe different phenomena characteristics of economic trends in situations compromised by the rapid spread of an epidemic, such as the unequal impact on the various wealth classes and the risk of a shrinking middle class.}
\\[+.2cm]

%\tableofcontents

\section{Introduction}

The rapid spreading of the COVID-19 epidemic in western countries and the consequent lockdown measures assumed by the governments to control  and limit its effects, will unequivocally lead to important consequences for their economies.  Clearly, a precise quantification of the damages of the disease spreading in wealth distribution is an extremely difficult problem which requires the knowledge of a large number of unknown variables and relationships among them. In an attempt to gain a better understanding of the mechanisms underlying the possible consequences of this new epidemic on Western economies, it is useful to rely on simplified mathematical models which, although based on a few evident universal characteristics, can be analysed to provide answers on possible scenarios. 

Driven by this objective,  in this paper we develop a mathematical framework to jointly model wealth distribution and spread of the infectious disease in a multi agent system,   by integrating the dynamics of a classical compartmental model in epidemiology \cite{CS78,HWH00,KM05} with the kinetic model of trading activity introduced in \cite{CPT05}.  While this paper concentrates on the classical SIR dynamics, it can be easily extended to incorporate other epidemiological dynamics, like the classical endemic models \cite{HWH00}, or to couple the SIR dynamics with other social and economical phenomena of multi agent systems \cite{PT13}. Resorting to SIR model means that we abstract away from disease related mortality. Clearly, this is a strong hypothesis, as it excludes demographic interaction. However, it is currently believed that the mortality rate of COVID-19 is low enough to justify this assumption.  In addition, the introduction of disease-related mortality introduces an additional state variable, population size, which makes the complete mathematical description of the economy at the end of the epidemic more complex. 

An easy way to understand epidemiology models is that they specify movements of individuals between different states based on some matching functions or laws of motion. According to the classical SIR models \cite{HWH00}, agents in the system are split into three classes,  the susceptible, who can contract the disease,  the infectious, who have already contracted the disease and can transmit it,  and the recovered, who are healed and immune or isolated. Consequently, with respect to the modeling assumptions made in \cite{CPT05}, the forthcoming kinetic model is composed by a system of three kinetic equations, each of one describing the time evolution of the wealth distribution density in the class, and taking into account both the movements of agents from one class to others, and the trading activity between agents of the same class or of different classes.

The elementary trading interactions are suitably modified to take into account the personal response of agents in the different classes to the presence of the infectious disease. In particular,  since  the saving  is  expected to change in response to disease incidence \cite{GLN14}, we will assume that agents in each class are represented by a different saving propensity. In addition, to take into account the variation of  the market  in presence of a reduced trading activity during the infectious disease, we introduce a variable in time variance of the random effects modeling the external market risks. Although the resulting model is particularly simplified compared to the complexity of the problem, its foundations are based on models that are currently references in their respective fields. From the epidemic side, SIR model is widely used in most applications (cf. \cite{HWH00} and the references therein). Also, from the side of statistical mechanics of multi-agent systems,  the kinetic model introduced in \cite{CPT05} has been shown to flexible with respect to different problems coming from economy. 

Mathematical modeling of wealth distribution has seen in recent years a marked development \cite{Ang, Ang1, Ch02, ChaCha00, ChChSt05, DY, DMT, DMT1, DS,GSV, SGD, Sl04,TTZ},  mainly linked to the understanding of the mechanisms responsible of the formation of Pareto tails \cite{Par, PT-ss}. However,  in most of the models considered so far,  the explicit form of the equilibrium density, which represents one of the main aspects linked to the validity of the model in its economic setting,  is not explicitly known,  and only few relatively simple models can be treated analytically \cite{BaTo, BaTo2,Kat}. 

For this reason, the description of the evolution of the personal wealth revealed to be successful by resorting to a description in terms of a Fokker--Planck type equation, that allows for an explicit computation of the steady state distribution.  In \cite{BM} Bouchaud and Mezard  introduced a simple model of economy, where the time evolution of wealth is described by an equation capturing both exchange between individuals and random speculative trading, in such a way that the fundamental symmetry of the economy under an arbitrary change of monetary units is insured.  A Fokker--Planck type model was then derived  through a mean field limit procedure, with a solution becoming in time a Pareto (power-law) type distribution.

The key features of the steady state solution of the Fokker--Planck equation is that it is given by the inverse  Gamma density
 \be\label{equi2}
f^\infty(w) =\frac{(\mu-1)^\mu}{\Gamma(\mu)}\frac{\exp\left(-\frac{\mu-1}{w}\right)}{w^{1+\mu}},
 \ee
 where $\mu >1$ is a positive constant which can be related to the interaction parameters. 
In agreement with the observations of the Italian economist Vilfredo Pareto \cite{Par} on the distribution of wealth, the equilibrium density \fer{equi2} exhibits a power-law tail for large values of the wealth  variable.  

The Fokker--Planck equation in \cite{BM} appears  as limit of different kinetic models. In particular, it was obtained  in \cite{CPT05} via an asymptotic procedure applied to a Boltzmann-type kinetic model for binary trading in presence of risks. Also, the same equation with a modified drift term appears when considering suitable asymptotics of Boltzmann-type equations for binary trading in presence of taxation, in the case in which taxation is described by redistribution operators introduced in \cite{BST, DPT}. Systems of Fokker--Planck equations  have been considered in \cite{DT2} to model wealth distribution in different countries which are coupled by mixed trading, and to study the evolution of wealth in a society with agents using personal knowledge to trade \cite{PT-k}. These results contributed to retain that the kinetic model in \cite{CPT05} and its Fokker--Planck asymptotics represent a quite satisfactory description of the time-evolution of wealth density towards a Pareto-type equilibrium in a trading society. 

While in normal trading activity it is commonly  assumed that a large part of agents behave in a similar way, in presence of  an extraordinary situation like the one due to the epidemic, it is highly reasonable to conjecture that the behavior of individuals is strictly affected by their personal situation in terms of health or wealth. 
In order to simplify the treatment, in this work we focus on the assumption that it is the state of health that changes people's behavior in the economic field. This assumption is justified both thinking that susceptible people understand that consuming and working less reduces the probability of becoming infected, and the country does not have a government funded health care system. More generally, the model can be clearly extended to consider more realistic dependencies on the wealth of individuals.  However, this does not change the essential conclusions of our analysis, namely that the spread of the disease seriously affects the distribution of wealth, which, unlike the situation in the absence of epidemics, can converge towards a stationary state with a bimodal form \cite{Gup, GCC16}. 

The rest of the paper is organized as follows. In Section \ref{model}, we shall introduce  the  system of three SIR-type kinetic equations combining wealth dynamics and spread of infectious disease in a multi agent interacting system. Then, the qualitative analysis of the system of kinetic equation is briefly presented in Section \ref{sec:analysis}. For a better reading, the mathematical details leading to the main results of this Section are postponed to the Appendix \ref{sec:appendix}. 
In Section \ref{FP-limit} we will show that in a suitable asymptotic procedure the solution to the kinetic system tends towards the solution of a system of three SIR-type Fokker-Planck type equations.  Once the  system of Fokker--Planck type equations  has been derived, we recover  in a simplified case the explicit steady state of the wealth distribution, which is found to have a bimodal shape.
In Section \ref{numerics} we will investigate at a numerical level the relationships  between the solutions of the kinetic system of Boltzmann type and its Fokker-Plack asymptotics. Simulations confirm the ability of the model to describe different phenomena characteristic of the economic trend in situations compromised by the rapid spread of an epidemic, like the unequal impact over the various wealth classes and the risk of a shrinking middle class.

\section{Wealth dynamics in epidemiologic models}\label{model}

The goal of this section is to build a kinetic system suitable to describe the evolution of wealth in a multi-agent system under the spread of an infectious disease.  As in classical SIR models \cite{HWH00}, the entire population is divided into three classes of agents: the susceptible, who can contract the disease; the infectious, who have already contracted the disease and can transmit it; and the recovered, who are healed and immune or isolated. 

Agents in the system are considered indistinguishable \cite{PT13}. This means that an agent's state at any instant of time $t\ge 0$ is completely characterized by the wealth $w \ge0$, measured in some unit.
We denote by $f_S(w,t)$, $f_I(w,t)$ and $f_R(w,t)$, $w\in \mathbb{R}_+$, the  wealth distributions at time $t > 0$ of susceptible, infectious and recovered individuals, respectively. The total wealth distribution is then recovered as
\[
f(w,t)=f_S(w,t)+f_I(w,t)+f_R(w,t).
\]
As outlined in the introduction, we do not introduce disease related mortality. Therefore,  we can fix the total wealth distribution to be a probability density
\[
\int_{\mathbb{R}_+} f(w,t)\,dw = 1,\quad t>0.
\]
As a consequence 
\begin{equation}
S(t)=\int_{\mathbb{R}^+}f_S(w,t)\,dw,\quad I(t)=\int_{\mathbb{R}^+}f_I(w,t)\,dw,\quad R(t)=\int_{\mathbb{R}^+}f_R(w,t)\,dw,
\end{equation}
denote the fractions of the population that are susceptible, infectious and recovered respectively. We also denote the total mean wealth as
\[
m(t)=\int_{\mathbb{R}^+}wf(w,t)\,dw,
\]
and the relative mean wealths as
\begin{eqnarray}
m_S(t)=\int_{\mathbb{R}^+}wf_S(w,t)\,dw,\quad m_I(t)=\int_{\mathbb{R}^+}w f_I(w,t)\,dw,\quad m_R(t)=\int_{\mathbb{R}^+}wf_R(w,t)\,dw.
\end{eqnarray}
In what follows, we assume that the evolution of the densities obeys the classical SIR model \cite{HWH00}, and that the various populations in the model act differently in the  economic process. The kinetic model then follows combining the epidemic process  with the wealth dynamics, as modeled by \cite{CPT05}.
This gives the system 
\begin{eqnarray}
\frac{\partial f_S(w,t)}{\partial t} &=& -K(w,t) f_S(w,t) + \sum_{J\in \{S,I,R\}} Q(f_S,f_J)(w,t)
\label{eq:SIRw1}\\
\frac{\partial f_I(w,t)}{\partial t} &=& K(w,t) f_S(w,t) - \gamma(w) f_I(w,t) + \sum_{J\in \{S,I,R\}} Q(f_I,f_J)(w,t)
\label{eq:SIRw2}\\
\frac{\partial f_R(w,t)}{\partial t} &=& \gamma(w) f_I(w,t) + \sum_{J\in \{S,I,R\}} Q(f_R,f_J)(w,t)\label{eq:SIRw3}
\end{eqnarray}
%\begin{eqnarray}
%\frac{\partial f_S(w,t)}{\partial t} &=& -K(w,t) f_S(w,t) + Q(f_S,f_S)(w,t) + Q(f_S,f_R)(w,t)
%\label{eq:SIRw1}\\
%\frac{\partial f_I(w,t)}{\partial t} &=& K(w,t) f_S(w,t) - \gamma(w) f_I(w,t)
%\label{eq:SIRw2}\\
%\frac{\partial f_R(w,t)}{\partial t} &=& \gamma(w) f_I(w,t) + Q(f_R,f_R)(w,t) + Q(f_R,f_S)(w,t)
%\label{eq:SIRw3}
%\end{eqnarray}
where $\gamma(w)$ is the recovery rate for people with wealth $w$, and the transmission of the infection is governed by the function
\be\label{eq:kappa}
K(w,t) = \int_{\mathbb{R}_+} \beta(w,w_*)f_I(w_*,t)\,dw_*,
\ee 
with $\beta(w,w_*)$ denoting the contact rate between people with wealths $w$ and $w_*$. 

In equations \eqref{eq:SIRw1}-\eqref{eq:SIRw3} the operator $Q$ characterizes the wealth evolution due to trading between agents of the same class, or between agents of different classes, and is built accordingly to the CPT model \cite{CPT05}. We adopted the notation $f_J$, $J\in \{S,I,R\}$ to denote the three different classes of individuals.
Clearly, the presence of the epidemic is responsible for suitable modifications that will be enlightened in details in the following.

Let us consider the wealth interactions in equation \eqref{eq:SIRw1} concerning the susceptible part of the population. Denoting with $w$ the wealth of the susceptible individuals and with $w_*$ the wealth of individuals belonging to the various classes of susceptible, infectious and recovered, the binary trades are characterized by
\begin{equation}
\begin{split}
w' &= (1-\a_S) w + \a_J w_* + \eta_{SJ} w\\[-.2cm]
&\hskip 6cm J\in \{S,I,R\}\\[-.2cm]
w_*' &= (1-\a_J) w_* + \a_S w + \tilde\eta_{SJ} w_*,
\end{split}
\label{eq:binS}
\end{equation}
where $\a_J\in (0,1)$ are transaction coefficients, while the market risk variables $\eta_{SJ}\ge -\a_S$ and $\tilde\eta_{SJ}\ge - \a_J$ are independent and identically distributed random variables with zero mean and the same time-dependent variance $\sigma^2(t)$ (since the risk in the market does not depend on the particular class of trading agents) that will be discussed later. 

Similarly we can consider interactions in equation \eqref{eq:SIRw2} concerning the infectious part of the population, that, denoting with $w$ the wealth of the infectious individuals, read 
\begin{equation}
\begin{split}
w' &= (1-\a_I) w + \a_J w_* + \eta_{IJ} w\\[-.2cm]
&\hskip 6cm J\in \{S,I,R\}\\[-.2cm]
w_*' &= (1-\a_J) w_* + \a_I w + \tilde\eta_{IJ} w_*,
\end{split}
\label{eq:binI}
\end{equation}
where now $\eta_{IJ}\ge -\a_I$ and $\tilde\eta_{IJ}\ge - \a_J$ are again random variables with zero mean and variance $\sigma^2(t)$.

Finally, concerning interactions in equation \eqref{eq:SIRw3} if $w$ is the wealth of the recovered individuals we get
\begin{equation}
\begin{split}
w' &= (1-\a_R) w + \a_J w_* + \eta_{RJ} w\\[-.2cm]
&\hskip 6cm J\in \{S,I,R\}\\[-.2cm]
w_*' &= (1-\a_J) w_* + \a_R w + \tilde\eta_{RJ} w_*,
\end{split}
\label{eq:binR}
\end{equation}
with $\eta_{RJ}\ge -\a_R$ and $\tilde\eta_{RJ}\ge - \a_J$ random variables with zero mean and  variance $\sigma^2(t)$.

%The binary trade between recovered follows the same rules, but with a different transaction coefficients $\a_R\in (0,1)$ in the form 
%\begin{equation}
%\begin{split}
%w' &= (1-\a_R) w + \a_R w_* + \eta_2 w\\
%w_*' &= (1-\a_R) w_* + \a_R w + \tilde\eta_2 w_*.
%\end{split}
%\label{eq:binR}
%\end{equation}
%We assume that the risk variables $\eta_2$, $\tilde\eta_2$ have zero mean and the same variance $\sigma^2(t)$, since the risk in the market does not depend on the particular class of trading agents. On the contrary we assume that the transaction parameters are different, with $\lambda_R > \lambda_S$. 
%Finally, mixed trading follows the rules 
%\begin{equation}
%\begin{split}
%v' &= (1-\a_S) v + \a_R w + \eta_3 v\\
%w' &= (1-\a_R) w + \a_S v + \tilde\eta_3 w.
%\end{split}
%\label{eq:binRS}
%\end{equation}

\subsection{Inside the binary trades}\label{sect:binary}

To clarify the modeling assumptions that lead to the present choice of trades, we recall the main consequences related to the original choice made in \cite{CPT05}.
The trade between agents have been modeled to include the idea that wealth changes hands for a specific reason: one agent intends to invest his wealth in some asset, property etc. in possession of his trade partner. Typically, such investments bear some speculative risk, and either provide the buyer with some additional wealth, or lead to the loss of wealth in a non-deterministic way. An easy realization of this idea consists in coupling the saving propensity parameter \cite{Ch02, ChaCha00} with some risky investment that yields an immediate gain or loss proportional to the current wealth of the investing agent. Hence $0 < \a_J < 1$, $j\in\{S,I,R\}$ are the parameters which identify the saving propensities $1-\lambda_J$, namely the intuitive behavior which prevents the agents to put in a single trade the whole amount of his money. The choice $ \lambda_R > \lambda_S$, for example, reflects the fact that susceptible individuals can be more cautious in the market and tend to save their wealth, since they understand that consuming and working less reduces the probability of infection \cite{macro}. On the other hand, infectious individual have limited possibilities to act on the market and, as we will see, asymptotically disappears from the wealth dynamic. 

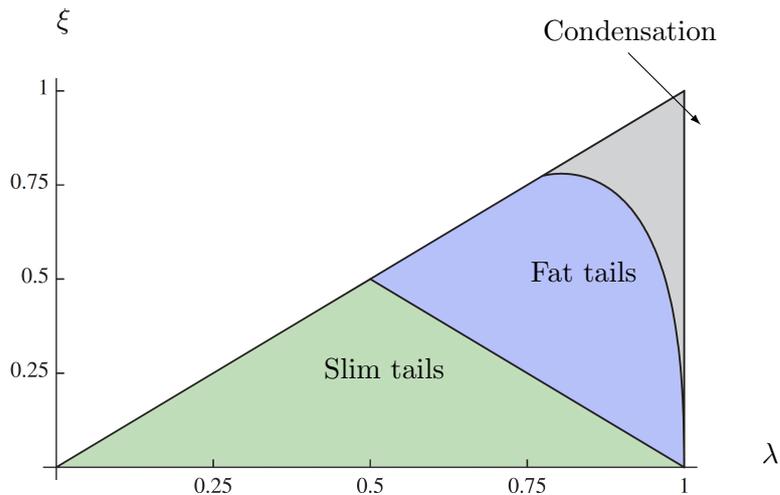
\begin{figure}[tb]
\begin{center}
\setlength{\unitlength}{0.1\textwidth}
\begin{picture}(9.8,4.5)
\put(1.5,0){\includegraphics[scale=0.2]{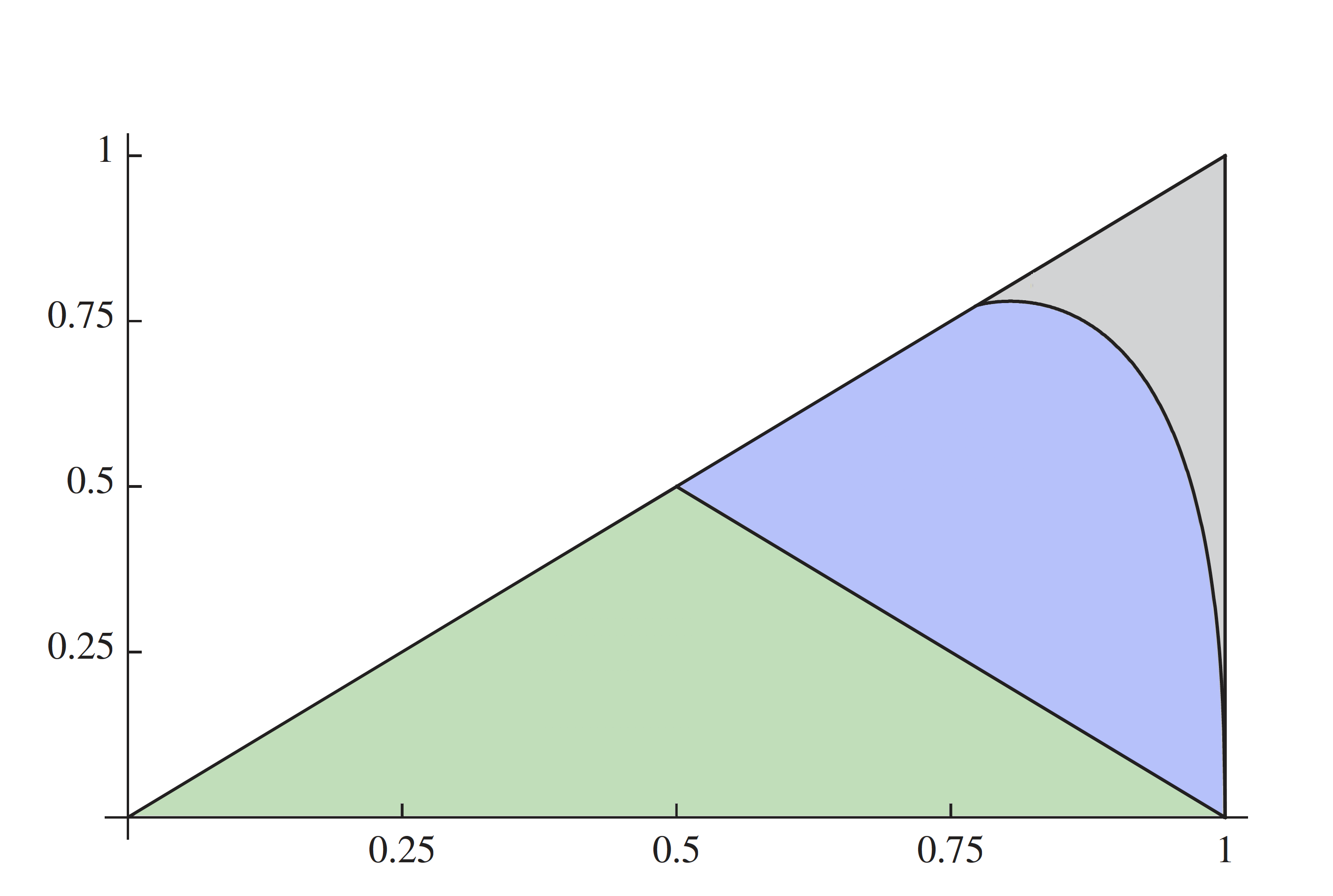}}
\put(4.3,1.1){{Slim tails}}
\put(6,1.9){{Fat tails}}
\put(6.1,3.9){{Condensation}}
\put(6.8,3.8){\vector(1,-1){.6}}
\put(2.1,4){$\xi$}
\put(7.9,0.4){$\lambda$}
\end{picture}
\end{center}
\caption{Parameter ranges with different tail regimes for homogeneous interactions}
\label{fg:pareto}
\end{figure}

The model describes two main types of wealth interactions. Interactions between individuals with the same status (homogeneous interactions) and interactions between individuals with different status (heterogeneous interactions). Consider the first case, for example, the wealth interaction between susceptible individuals. The role of the risk variables $\eta_{SS}$ and $\tilde\eta_{SS}$, for $J=S$ in \eqref{eq:binS} have been clarified in \cite{MT08} resorting to a specific choice. The easiest one leading to interesting results is $\eta_{SS}= \pm \xi$ where each sign comes with probability $1/2$.   The factor $\xi \in(0, \a_S)$, should be understood as the intrinsic risk of the market: it quantifies the fraction of wealth agents are willing to 'gamble' on. Within this choice, one can display the various regimes for the steady state of wealth in dependence of $\lambda=\a_S$ and $\xi$ (see Figure \ref{fg:pareto}). In the zone corresponding to low market risk, the wealth distribution shows a 'socialistic' behavior with slim tails. Increasing the risk, one falls into a 'capitalistic' zone, where the wealth distribution displays a Pareto fat tail. A minimum of saving ($\a > 1/2$) is necessary for this passage; this is expected since if wealth is spent too quickly after earning, agents cannot accumulate enough to become rich. Inside the capitalistic zone, the Pareto index decreases from $+\infty$ at the border with the socialist zone to unity. Finally, one can obtain a steady wealth distribution which is a Dirac delta located at zero. Both risk and saving propensity are so high that a marginal number of individuals manages to monopolize all of the society's wealth. 

On the other hand, the interaction between individuals with different status presents similar dynamics but with an important difference. The role of the risk variable, as we have said, is the same, since it represents a common factor in the economic system, but the different individual saving propensities lead to a different exchange of wealth between agents. Let us consider, for example, the interaction between susceptible and recovered individuals in \eqref{eq:binS} for $J=R$. Compared to the previous dynamics between only susceptible agents, if we assume $\lambda_R = \lambda_S+\delta$, $\delta >0$ it is evident that the risk of loss of recovered becomes $\delta +\tilde\eta_{SR}$ and has no more zero average. A similar but reversed argument shows that this produces a systematic gain for the susceptible.

In the view of the previous discussion, it is reasonable to assume that in the presence of a significant spread of the epidemic, the risk variance tends to increase. This is in agreement, for example, to the market reactions we observed during the COVID-19 spreading at the announcements of the new numbers of infectious people in the various countries \cite{ZHJ}. A simple example is based on assuming that the value takes into account the instantaneous number of infected people
 \be\label{eq:sigma}
  \sigma (t) = \sigma \left(1 + \alpha I(t)\right).
 \ee 
Possible long term memory effects of the infection-dependent risk variance can be suitably introduced. We postpone a more detailed comparison to the numerical examples  presented in the last section of the paper.
% \be\label{eq:sigma}
%  \sigma (t) = \sigma \left(1 + \alpha \int_0^t I(s)\,ds \right)^{-1}.
% \ee
% where $0 <\alpha $ is a suitable parameter which quantifies the importance given to  the spreading disease by the society. Other choices, of course, are possible we refer to the last Section containing various numerical simulations for further examples.

  %For example, $\alpha = 0$ in the case in which the society is concerned with the seasonal flu.
 
 We are now ready to define the operator $Q(\cdot,\cdot)$ in the various cases. As observed in \cite{CPT05, PT13} a convenient way to express the operator is based on its weak form, namely the way the operator acts on observables. Let $\phi(v)$ denote an observable function and let us define with $\langle \cdot \rangle$ the expected value with respect to the pair $\eta_i$, $\tilde\eta_i$ in the interaction process. Then, adopting the previous notations for the wealth of susceptible and recovered, in \eqref{eq:SIRw1} the action of $Q(f_S,f_J)(w,t)$ on $\phi(w)$ is given by

%In strong form the operator has the general structure 
%\begin{equation}
%Q(F,G)=\int_{\mathbb{R}^2\times\mathbb{R}_+} \left\{\beta(w,w_*)J F('w,t)G('w_*,t)-\beta(w',w'_*)F(w,t)G(w_*,t)\right\}\,dw_*\,d\eta\,d\eta_*,
%\label{eq:econ}
%\end{equation}
%where $('w,'w_*)$ are the pre-interaction wealths that generate the pair $(w,w_*)$. In \eqref{eq:econ} $J$ is Jacobian of the transformation of $(w, w_*)$ into $(w', w'_*)$ and the kernel $\beta$ is related to the details of the binary interaction
%\begin{equation}
%\beta(w',w'_*)=\mu(\eta)\mu(\eta_*)\Psi(w' \geq 0)\Psi(w'_*\geq 0),
%\label{eq:ker}
%\end{equation}
%with $\Psi(\cdot)$ the indicator function that will not allow agents to have negative wealth.  

%The binary interaction rules in \eqref{eq:econ} have the form
%\begin{equation}
%\begin{split}
%w' &= (1-\alpha) w + \alpha w_* + \eta w\\
%w_*' &= (1-\alpha) w_* + \alpha w + \eta_* w,
%\end{split}
%\label{eq:bin}
%\end{equation}
%where $\alpha\in (0,1)$ is the transaction coefficient, while $\eta$ and $\eta_*$ are random variables with the same distribution $\mu(\eta)$ with variance $\sigma^2$ and zero mean. The first term in \eqref{eq:bin} is related to the marginal saving propensity of the agents, the second corresponds to the wealth transaction, and the last contains the effects of an open economy describing the market returns. Note that, the zero debts condition 
%in \eqref{eq:ker} implies that the fluctuations induced by $\eta$ and $\eta_*$ produce a grow of the overall wealth.
%
%The binary interaction operator admits a simpler representation in weaker form
\begin{equation}
\int_{\mathbb{R}_+}Q(f_S,f_J)\phi(w)\,dw=\Big\langle\int_{\mathbb{R}^2_+}  f_S(w,t)f_J(w_*,t)(\phi(w')-\phi(w))\,dw_*\,dw\Big\rangle,
\label{eq:econS}
\end{equation}
where $w'$ is defined by \eqref{eq:binS}. Similarly $Q(f_I,f_J)(w,t)$ in \eqref{eq:SIRw2} is characterized by
\begin{equation}
\int_{\mathbb{R}_+}Q(f_I,f_J)\phi(w)\,dw=\Big\langle\int_{\mathbb{R}^2_+}  f_I(w,t)f_J(w_*,t)(\phi(w')-\phi(w))\,dw_*\,dw\Big\rangle,
\label{eq:econI}
\end{equation}
with $w'$ given by \eqref{eq:binI} and $Q(f_R,f_J)(w,t)$ in in \eqref{eq:SIRw3} by
\begin{equation}
\int_{\mathbb{R}_+}Q(f_R,f_J)\phi(w)\,dw=\Big\langle\int_{\mathbb{R}^2_+}  f_R(w,t)f_J(w_*,t)(\phi(w')-\phi(w))\,dw_*\,dv\Big\rangle,
\label{eq:econR}
\end{equation}
being $w'$ defined as in \eqref{eq:binR}

A couple of remarks are important before moving on to a more detailed analysis of the model.

\begin{remark}~ 
\begin{itemize}
\item 
For simplicity, we assumed the parameters characterizing the saving propensities constant. More generally it is possible to consider $\lambda_J=\lambda_J(I)$ functions of the number of infected people as it is natural that individuals take into account not only their condition but also the epidemic as a whole. For example, by determining the functions $\lambda_J(I)$ on the basis of suitable utility functions \cite{macro}. We are aware that this would make the model more realistic, but since the essential characteristics of the dynamics are not altered and on the contrary the mathematical analysis of the model would become extremely difficult, in the following we will focus on mathematically treatable scenarios that allow us to draw general  conclusions.  
\item A general criticism may concern the assumption that total wealth is preserved in the exchange processes \cite{GKLO}. However, it has been observed, see \cite{PT-ss}, that non-conservative models can be reformulated in conservative form under an appropriate self-similar scaling that normalizes the overall wealth. In this sense, the original wealth model \cite{CPT05} is not conservative at the binary trade level since it allows the random stochastic multiplicative process to produce wealth, therefore taking into account not only exchanges but also production. However, its asymptotic analysis in a conservative context makes it possible to identify the observed stylized facts of a real economy, such as the formation of power law tails in the wealth distribution \cite{PT-ss, MT08}. 
\item In \fer{eq:econS}-\fer{eq:econR}  we assumed an identical interaction frequency (equal to unity). This is clearly a simplified but reasonable choice, which is based on the assumption that the agent entering a binary trade is generally not aware of the health status of the other agent in relation to the epidemic. However, both the quantitative and qualitative analysis of the system can easily be generalised to operators with different weights.
\end{itemize}
\end{remark} 

\subsection{Properties of the kinetic model}\label{sec:analysis}
This Section deals with the mathematical analysis of  the Boltzmann type system \eqref{eq:SIRw1}-\eqref{eq:SIRw3}, with the principal objective to describe the large-time behavior of the solution, its convergence to a unique final state, as well as its main properties. We will present only the main results, by postponing the details of the computations in Appendix \ref{sec:appendix}. To avoid inessential difficulties in the computations, we will limit ourselves to consider a the simplified version. We assume  the interaction rate of infections $\beta(w,w_*)=\beta$ constant  in expression \fer{eq:kappa},  a constant function $\gamma(w) = \gamma$,  and $\alpha =0$ in \fer{eq:sigma}, so that $\sigma(t) =\sigma$. In this case, the system \eqref{eq:SIRw1}-\eqref{eq:SIRw3}SIR  can be fruitfully rewritten resorting to Fourier transforms. Note however that we can resort to Fourier transform also in the more realistic case in which the function $K(w,t)$ in \fer{eq:kappa} is given by a convolution, that is
\be\label{eq:conv} 
K(w,t) = \int_{\mathbb{R}_+} \beta(w-w_*)f_I(w_*,t)\,dw_*,
\ee 
with $\beta (w) \in L_1(\R_+)$. 

Given a function $f(w)\in L_1(\R_+)$, we define its Fourier transform by
\[
\hat{f}(\xi) = \int_{\R} e^{-iw \xi}\,f(w)\, dw.
\]
Then the system \eqref{eq:SIRw1}-\eqref{eq:SIRw3} takes the form
\begin{eqnarray}
\frac{\partial \fS(\xi,t)}{\partial t} &=& -\beta I(t) \fS(\xi,t) + \sum_{J\in \{S,I,R\}} \hat Q(\fS,\fJ)(\xi,t)
\label{eq:SIRf1}\\
\frac{\partial \fI(\xi,t)}{\partial t} &=& \beta I(t) \fS(\xi,t) - \gamma \fI(\xi,t) + \sum_{J\in \{S,I,R\}} \hat Q(\fI,\fJ)(\xi,t)
\label{eq:SIRf2}\\
\frac{\partial \fR(\xi,t)}{\partial t} &=& \gamma \fI(\xi,t) + \sum_{J\in \{S,I,R\}} \hat Q(\fR,\fJ)(\xi,t)\label{eq:SIRf3}
\end{eqnarray}
The operators  $ \hat Q(\fH,\fJ)(\xi,t)$,  are easily defined in terms of the Fourier transforms of  $\fJ$, with ${J, \in \{S,I,R\}}$ by choosing in \fer{eq:econS}- \fer{eq:econR} as observable function $\phi(w) = \exp\{ -iw\xi\}$ \cite{DMT, PT13}. One easily obtains
\begin{equation*}\label{QS}
\hat Q(\fH,\fJ)(\xi,t) = \langle\fH (A_{HJ}\xi, t)\rangle \fJ(\lambda_J\xi,t) -I_J(t) \fH(\xi,t).
\end{equation*}
where  $A_{HJ}$, with ${H,J \in \{S,I,R\}}$ denote the random parameters  
\be\label{eq:breve}
A_{HJ} = 1 -\lambda_H+\eta_{HJ}.
\ee
In what follows, let us suppose that these parameters satisfy the condition
\be\label{eq:regola}
\nu =  \max_ {H,J\in \{S,I,R\}}\big[ \lambda_J^2 + \langle A_{JH}^2 \rangle \big] < 1.
\ee
By evaluating the mean values of the random parameters $A_{JH}^2$ it can be shown that this condition is verified at any time such that
\be\label{eq:strong}
\sigma < 2\min_{H\in \{S,I,R\}}\lambda_H(1-\lambda_H).
\ee
Inequality \fer{eq:strong} establishes a relationship between  the saving propensities and the risks of the market. We anticipate that, in the scaling \fer{eq:quasi} leading to the Fokker--Planck description presented in Section \ref{FP-limit},  condition \fer{eq:strong}  for $\varepsilon \to 0 $  reads
\be\label{eq:strong1}
\min_{H\in \{S,I,R\}}\frac{2 \lambda_H}\sigma >1,
\ee
which is the natural condition, in the simpler case of only one group of individuals, considered in \cite{BM, CPP}. 
The large-time behavior of solutions to systems like \fer{eq:SIRf1} - \fer{eq:SIRf3} can be studied by resorting to a class of metrics which has been shown to be particularly suitable for bilinear equations of Boltzmann type with Maxwell interactions \cite{PT13}. Let $f$ ang $g$ be probability densities. Then, for a given constant $s>0$ we define 
\be\label{eq:ds}
d_s(f,g) = \sup_{\xi \in \R} \frac{|\ff(\xi) -\fg(\xi)|}{|\xi|^s}
\ee
The metric $d_s$ is finite any time the densities $f$ and $g$ possess equal moments up to $[s]$, the entire part of $s$, and up to $s-1$ if $s\in \mathbb N$.

The following theorem,  will certify that the kinetic system  \eqref{eq:SIRw1}-\eqref{eq:SIRw3} is well posed from a mathematical point of view. 

\begin{thm}\label{thm:main}
	Let $f_J(w,t)$ and $g_J(w,t)$, $J\in  \{S,I,R\}$ be two solutions of the kinetic system \eqref{eq:SIRw1}-\eqref{eq:SIRw3}, corresponding to initial values $f_J(w,0)$ and $g_J(w,0)$ such that $d_2(f_J(w,0), g_J(w,0))$,  $J\in  \{S,I,R\}$, is finite. Then, provided condition \fer{eq:regola} holds, the Fourier based distance $d_2(f_J(w,t), g_J(w,t))$ decays exponentially in time to zero, and the following holds
	\be\label{eq:dec}
	\sum_{J\in  \{S,I,R\}} d_2(f_J(w,t), g_J(w,t)) \le \sum_{J\in  \{S,I,R\}} d_2(f_J(w,0), g_J(w,0))\exp\{(-(1-\nu)t\}.
	\ee
\end{thm}
The details of the proof of this result are presented in Appendix \ref{sec:appendix}. 

In view of the completeness of the $d_s$ metrics for $1<s<2$, see \cite{BG}, and since convergence in $d_2$-metric implies convergence in $d_s$-metric for any $s <2$, see \cite{PT13}, the existence of the long-time limit $f_J^\infty(w)$, $J\in \{S,I,R\}$ can be concluded directly from the contractivity of the system in $d_2$ metric.

Theorem \ref{thm:main} allows to further investigate properties of the limit densities $f_J^\infty(w)$, $J\in \{S,I,R\}$. To this aim, we use the fact that the relative mass densities the Boltzmann type system \eqref{eq:SIRw1}-\eqref{eq:SIRw3} satisfy the classical SIR model
\begin{eqnarray*}
\frac{d S(t)}{d t} &=& -\beta I(t) S(t) 
\label{eq:SIR10}\\
\frac{d I(t)}{d t} &=& +\beta I(t) S(t) - \gamma I(t)
\label{eq:SIR20}\\
\frac{d R(t)}{d t} &=& \gamma I(t). 
\label{eq:SIR30}
\end{eqnarray*}
In this case, (cf.  \cite{HWH00} ) it is known  that $I(t)\to 0$, while $S(t)\to S^\infty \in [0,\gamma/\beta]$ solution of  
\[
I(0)+S(0)-S^\infty+\frac{\gamma}{\beta}\log\left(\frac{S^\infty}{S(0)}\right)=0.
\]
Therefore, if the initial densities $f_J(w, t=0)$,  $J\in \{S,I,R\}$ are such that
\be\label{eq:costa}
S( t=0) = S^\infty, \quad I( t=0) = 0, \quad R(t=0) = 1 - S^\infty,  
\ee
at any subsequent time $t>0$  the masses of the functions $f_J(w, t)$ still satisfy conditions \fer{eq:costa}. This implies that the limit density  $f_I^\infty(w)= 0$, while $f_J^\infty(w)$, $J\in \{S,R\}$ are steady solutions of the Boltzmann system
\begin{eqnarray*}\label{eq:staz}
\sum_{J\in \{S,R\}} Q(f_S,f_J)(w,t) =0 
\label{eq:SIRs1}\\
\sum_{J\in \{S,R\}} Q(f_R,f_J)(w,t) = 0 \label{eq:SIRs2}
\end{eqnarray*}
Indeed, denote by $f_J^\infty(t)$ $J\in \{S,I,R\}$ the solution to the system  \eqref{eq:SIRw1}-\eqref{eq:SIRw3} with initial datum $f_J^\infty$ $J\in \{S,I,R\}$, thus satisfying \fer{eq:costa}. Then, for any other density $f_J(w, t)$,  $J\in \{S,I,R\}$ satisfying the same conditions \fer{eq:costa}, Theorem \ref{thm:main} implies
\[
d_1(f_J^\infty(t), f_J^\infty) \le  d_1[f_J^\infty(t), f_J(t+T)]  + d_1[f_J(t+T), f_J^\infty] \le
\]
\[
2^{3/2}\exp\left\{-\frac{(1-\nu)t}2\right\} d_1[f_J^\infty(t), f_J(t)] +  d_1[f_J(t+T), f_J^\infty] .
\]
The first bound follows from there interpolation property of the metric $d_s$ (cf. Proposition 2.1 in \cite{PT13}). The last expression can be made arbitrarily small by choosing $T$ large enough, so that $f_J^\infty(t) = f_J^\infty$ for all $T \ge 0$. In fact, the functions
$f_J^\infty$,  $J\in \{S,I,R\}$ are the {\em only} steady state with  values of the
masses given by \fer{eq:costa}; if $g_J^\infty$ is another steady state with the same
masses, then $d_1[f_J^\infty,g_J^\infty]$ is finite, and so, invoking
Theorem~\ref{thm:main} again,
\begin{align*}
d_1[f_J^\infty,g_J^\infty] &\leq e^{-r}d_1[f_J^\infty,g_J^\infty],
\end{align*}
which forces $f_J^\infty=g_J^\infty$.

\section{Fokker-Planck scaling and steady states}\label{FP-limit}
To analyze the asymptotic behavior of the model it is useful to resort to the so-called quasi-invariant trading limit which permits to derive the corresponding Fokker-Planck description to \eqref{eq:SIRw1}-\eqref{eq:SIRw2}. Similar asymptotic analysis was performed in \cite{CPP,DMT} for  a kinetic model for the distribution of wealth in a simple market economy subject to microscopic binary trades in presence of risk, showing formation of steady states with Pareto tails, in \cite{TBD} on kinetic equations for price formation, and in \cite{To1} in the context of opinion formation in presence of self-thinking. A general view about this asymptotic passage from  kinetic equations based on general interactions  towards Fokker--Planck type equations can be found in \cite{FPTT16}. Other relationships of this asymptotic procedure with the classical problem of the \emph{grazing collision limit} of the Boltzmann equation in kinetic theory of rarefied gases have been recently enlightened in \cite{GT-ec}.

Let us consider the case in which the trading produces only small modification of the wealth, and simultaneously the frequency of trading is increased. Following \cite{CPT05, FPTT16, PT13}, we scale the binary trades accordingly to
\begin{equation}\label{eq:quasi}
\lambda_S \to \varepsilon \lambda_S,\quad \lambda_I \to \varepsilon \lambda_I, \quad \lambda_R \to \varepsilon \lambda_R, \quad \sigma \to \sqrt{\varepsilon} \sigma,
\end{equation}
and similarly the functions governing the spread of the disease
\begin{equation}\label{eq:quasi_epi}
\beta(w,w_*) \to \varepsilon \beta(w,w_*),\quad \gamma(w) \to \varepsilon \gamma(w), 
\end{equation}
and denote with $Q_{\varepsilon}(\cdot,\cdot)$ the corresponding scaled interaction terms.

The limit procedure induced by the above scaling corresponds
to the situation in which are prevalent the
exchanges of wealth which produce an extremely small
modification the pre-interaction wealths (quasi-invariant interactions), but we are waiting enough time to still see the effects.

In fact, rescaling time as $t \to t/\varepsilon$, for small values of $\varepsilon$, and avoiding the time dependence on mean values and variance, we obtain (see \cite{CPT05, FPTT16, PT13} for details) for $J\in\{S,I,R\}$
\[
\frac1{\varepsilon}\int_{\mathbb{R}_+}Q_{\varepsilon}(f_S,f_J)\phi(w)\,dw = \int_{\mathbb{R}_+} \left\{-\phi'(w)\left(w\lambda_S J-m_J\lambda_J\right)+\frac{\sigma}{2}\phi''(w)w^2 J\right\}f_S(w,t)\,dv+O(\varepsilon),
\]
\[
\frac1{\varepsilon}\int_{\mathbb{R}_+}Q_{\varepsilon}(f_I,f_J)\phi(w)\,dw = \int_{\mathbb{R}_+} \left\{-\phi'(w)\left(w\lambda_I J-m_J\lambda_J\right)+\frac{\sigma}{2}\phi''(w)w^2 J\right\}f_I(w,t)\,dv+O(\varepsilon),
\]
\[
\frac1{\varepsilon}\int_{\mathbb{R}_+}Q_{\varepsilon}(f_R,f_J)\phi(w)\,dw = \int_{\mathbb{R}_+} \left\{-\phi'(w)\left(w\lambda_R J-m_J\lambda_J\right)+\frac{\sigma}{2}\phi''(w)w^2 J\right\}f_R(w,t)\,dv+O(\varepsilon).
\]
Reverting back to the original notations for the wealth as in \eqref{eq:SIRw1}-\eqref{eq:SIRw3}, the above equations, as $\varepsilon\to 0$, correspond to the following Fokker-Planck system in strong form
\begin{eqnarray}
&&\hskip -.8cm \frac{\partial f_S(w,t)}{\partial t} = -K(w,t) f_S(w,t) + \frac{\partial}{\partial w}\left[\left(w\lambda_S-\bar{m}(t)\right)f_S(w,t)\right] +\frac{\sigma(t)}{2} \frac{\partial^2}{\partial^2 w} (w^2 f_S(w,t)) 
\label{eq:SIRw1f}\\
\nonumber
&&\hskip -.8cm\frac{\partial f_I(w,t)}{\partial t} = K(w,t) f_S(w,t) - \gamma(w) f_I(w,t) + \frac{\partial}{\partial w}\left[\left(w\lambda_I-\bar{m}(t)\right)f_I(w,t)\right] \label{eq:SIRw2fa}\\[-.1cm]
\\[-.2cm]
\nonumber
&&\hskip .8cm +\frac{\sigma(t)}{2} \frac{\partial^2}{\partial^2 w} (w^2 f_I(w,t))
\label{eq:SIRw2f}\\
&&\hskip -.8cm\frac{\partial f_R(w,t)}{\partial t} = \gamma(w) f_I(w,t) + \frac{\partial}{\partial w}\left[\left(w\lambda_R-\bar{m}(t)\right)f_R(w,t)\right]+\frac{\sigma(t)}{2} \frac{\partial^2}{\partial^2 w} (w^2 f_R(w,t))
\label{eq:SIRw3f}
\end{eqnarray}
where
\begin{equation*}
\bar{m}(t)=\lambda_S m_S(t)+\lambda_I m_I(t)+\lambda_R m_R(t).
\end{equation*}
The above Fokker-Planck system is complemented with the boundary conditions at $w=0$ given by
\begin{equation*}
\frac{\partial}{\partial w} (w^2 f_J(w,t))\Big|_{w=0} = 0,\qquad f_J(0,t)=0,\quad J\in\{S,I,R\}.
\end{equation*}
Clearly, the steady state wealth distributions satisfy the ordinary differential equations corresponding to the equations \eqref{eq:SIRw1f}-\eqref{eq:SIRw3f} with the time derivatives set equal to zero. We discuss in the next Section a case where we can explicitly solve the steady state equations.

\subsection{An explicitly solvable case}\label{sect:explicit}

We verify in a simplified case, that the Fokker--Planck system \eqref{eq:SIRw1f}-\eqref{eq:SIRw3f}possesses an explicitly computable steady state, which can present a bimodal shape. Bimodal shapes are typical of situations of high stress in economy, and are investigated starting from the Argentinian crisis of the first year of the new century \cite{Gup}. This example also shows that a similar behavior can be expected in reason of the epidemic spreading.

Suppose that the interaction rate of infections $\beta(w,w_*)=\beta$ is constant, and $\alpha =0$ in \fer{eq:sigma}, so that $\sigma(t) =\sigma$. Then, integrating with respect to the wealth variable, thanks to conservation of the total wealth, we obtain that the relative mass densities satisfy the classical SIR model
\begin{eqnarray}
\frac{d S(t)}{d t} &=& -\beta I(t) S(t) 
\label{eq:SIR1}\\
\frac{d I(t)}{d t} &=& +\beta I(t) S(t) - \gamma I(t)
\label{eq:SIR2}\\
\frac{d R(t)}{d t} &=& \gamma I(t). 
\label{eq:SIR3}
\end{eqnarray}
In this case, it is known that $I(t)\to 0$, while $S(t)\to S^\infty \in [0,\gamma/\beta]$ solution of \cite{HWH00}   
\[
I(0)+S(0)-S^\infty+\frac{\gamma}{\beta}\log\left(\frac{S^\infty}{S(0)}\right)=0.
\]
Likewise, the system for the mean values read
\begin{eqnarray}
\frac{d m_S(t)}{d t} &=& -\beta I(t) m_S(t)+ S(t) \bar{m}(t) - \lambda_S m_S(t)
\label{eq:SIRm1}\\
\frac{d m_I(t)}{d t} &=& \beta I(t) m_S(t) - \gamma m_I(t)+ I(t) \bar{m}(t) - \lambda_I m_I(t)
\label{eq:SIRm2}\\
\frac{d m_R(t)}{d t} &=& \gamma m_I(t)+R(t) \bar{m}(t) - \lambda_R m_R(t). 
\label{eq:SIRm3}
\end{eqnarray}
Since $I(t)\to 0$, $m_I(t)\to 0$,   $m_S(t)\to m_S^\infty$ and $m_R(t)\to m_R^\infty$, where the asymptotic values of the means satisfy
\[
\lambda_R \frac{m^\infty_R}{R^\infty} = \lambda_S \frac{m^\infty_S}{S^\infty}, 
\]
together with the constraint $m^\infty_R+m^\infty_S=m$ by conservation of the total mean wealth. This gives the asymptotic values
\begin{equation}
m^\infty_S = \frac{\lambda_R S^\infty}{\lambda_R S^\infty+\lambda_S R^\infty} m,\qquad m^\infty_R = \frac{\lambda_S R^\infty}{\lambda_R S^\infty+\lambda_S R^\infty} m.
\label{eq:mean}
\end{equation}
From the above computations, formally as $t \to \infty$ in the Fokker-Planck system \eqref{eq:SIRw1f}-\eqref{eq:SIRw3f} we get that the stationary states $f^\infty_S(w)$ and $f^\infty_R(w)$ satisfy
\begin{eqnarray*}
\lambda_S\frac{\partial}{\partial w}\left[\left(w-\frac{m_S^\infty}{S^\infty}\right)f^\infty_S(w)\right] +\frac{\sigma}{2} \frac{\partial^2}{\partial^2 w} (w^2 f_S^\infty(w)) = 0\\
\lambda_R\frac{\partial}{\partial w}\left[\left(w-\frac{m_R^\infty}{R^\infty}\right)f^\infty_R(w)\right] +\frac{\sigma}{2} \frac{\partial^2}{\partial^2 w} (w^2 f_R^\infty(w)) = 0.
\end{eqnarray*}
Hence the steady states are explicitly computed as two inverse Gamma densities \cite{PT13}
\begin{equation}
f^\infty_S(w)=S^\infty \frac{\kappa^{\mu_S}}{\Gamma(\mu_S)}\frac{e^{-\frac{\kappa}{w}}}{w^{1+\mu_S}}, \qquad f^\infty_R(w)=R^\infty\frac{\kappa^{\mu_R}}{\Gamma(\mu_R)}\frac{e^{-\frac{\kappa}{w}}}{w^{1+\mu_R}}
\label{eq:fRS}
\end{equation}
with
\begin{equation}
\begin{split}
&\mu_S=1+2\frac{\lambda_S}{\sigma},\qquad\qquad \mu_R=1+2\frac{\lambda_R}{\sigma},\\
&\kappa = {(\mu_S-1)}\frac{m_S^\infty}{S^\infty} = {(\mu_R-1)}\frac{m_R^\infty}{R^\infty} = \frac{2\lambda_R\lambda_S}{\sigma(\lambda_R S^\infty+\lambda_S R^\infty)}m.
\label{eq:mu}
 \end{split}
\end{equation}

\begin{figure}
\begin{center}
\includegraphics[scale=0.42]{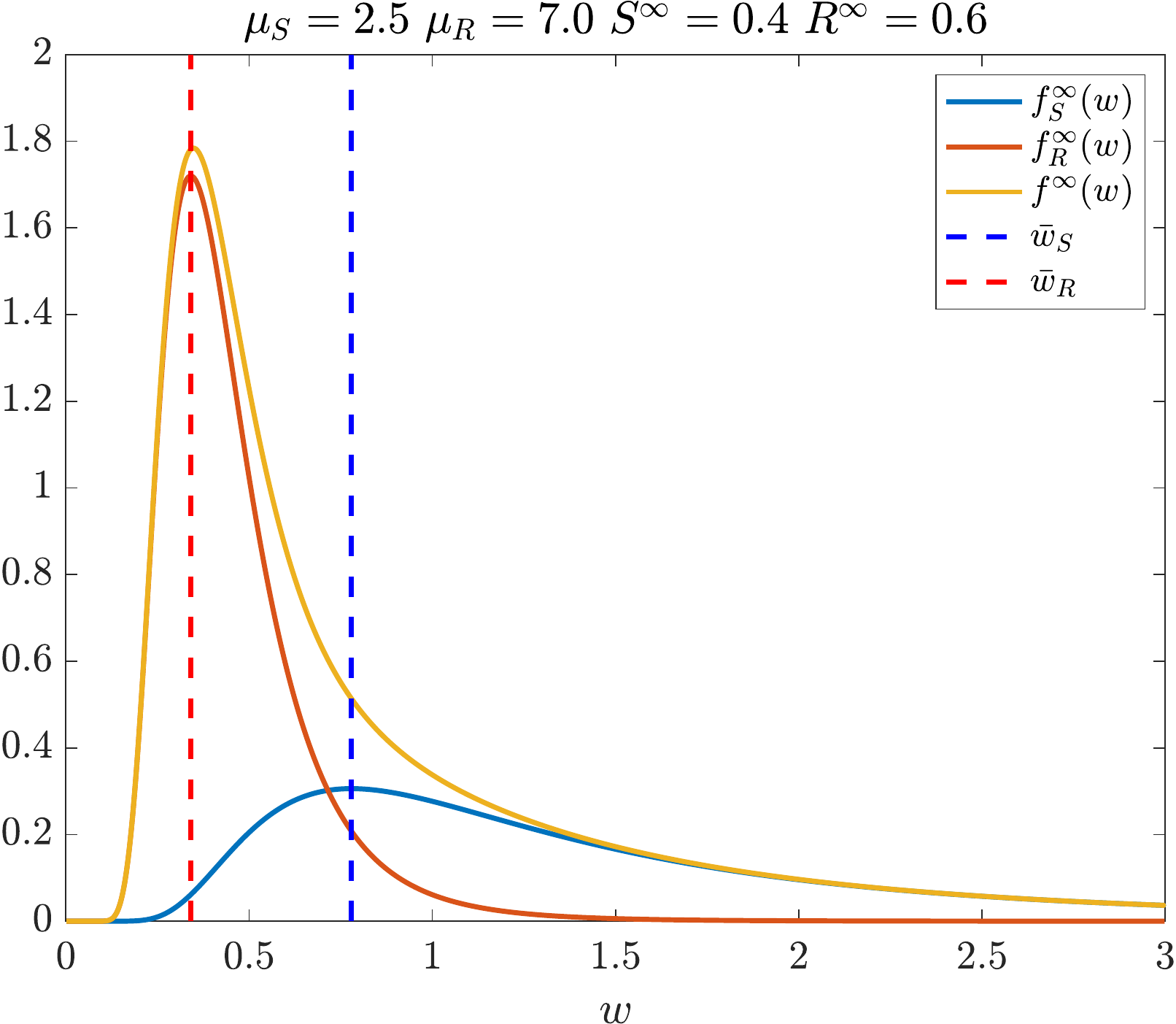}
\includegraphics[scale=0.42]{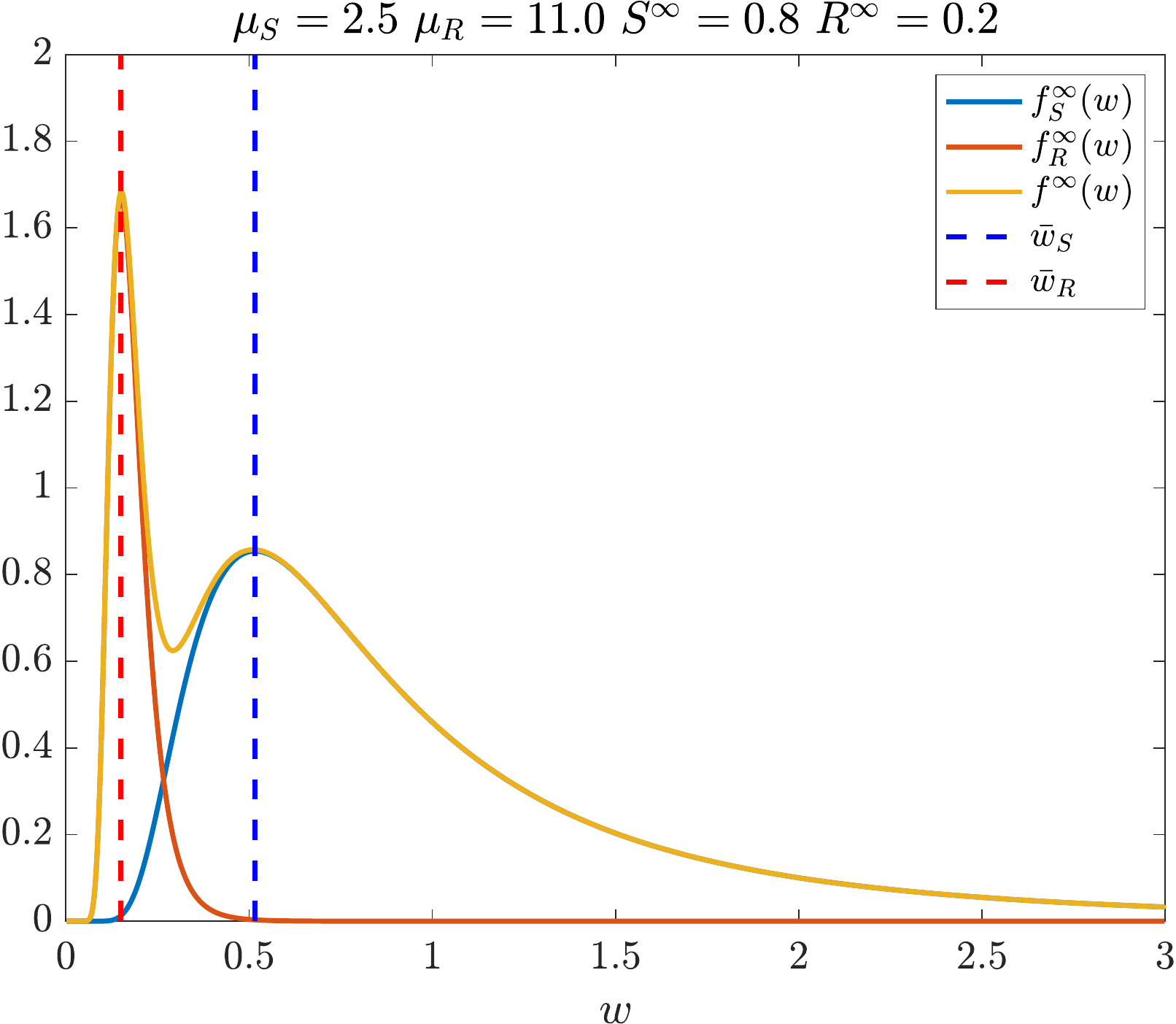}
\end{center}
\caption{Exact solutions for wealth distributions at the end of the epidemic \eqref{eq:fRS} in the Fokker-Planck approximation \eqref{eq:SIRw1f}-\eqref{eq:SIRw3f} for various choices of $\mu_S<\mu_R$, $S^\infty$ and $R^\infty$.}
\label{fg:bimodal}
\end{figure}

The  details of the trading activity at the basis of the kinetic description allow to characterize the tails of the distributions from \eqref{eq:mu}. Hence, a low value of the Pareto index is obtained in presence of small values of the parameter $\lambda_S$, $\lambda_R$  (small saving propensity of agents), or to high values of the parameter $\sigma$ (highly risky market).  
Therefore, the asymptotic wealth distribution is the mixture of two inverse Gamma densities of mass $S^\infty$ and $R^\infty$ respectively
\begin{equation}\label{eq:finf}
f^\infty(w)=f^\infty_S(w)+f^\infty_R(w),
\end{equation}  
with asymptotic means \eqref{eq:mean} and variances given by 
\[
{\rm Var}^\infty_S = \frac{\kappa^2}{(\mu_S-1)(\mu_S-2)},\qquad {\rm Var}^\infty_R = \frac{\kappa^2}{(\mu_R-1)(\mu_S-2)},\qquad \mu_R,\mu_S > 2.
\]
As a consequence, the wealth distribution has a bimodal structure, since the maximum of $f^\infty_S(w)$ and $f^\infty_R(w)$ are achieved, respectively, at the points 
\begin{equation}\label{eq:wbar_SR}
\begin{split}
\bar{w}_S &= \frac{\kappa}{\mu_S+1}=\frac{\lambda_R\lambda_S}{(\lambda_S + \sigma)(\lambda_R S^\infty+\lambda_S R^\infty)}m,\\
\bar{w}_R &= \frac{\kappa}{\mu_R+1}=\frac{\lambda_R\lambda_S}{(\lambda_R + \sigma)(\lambda_R S^\infty+\lambda_S R^\infty)}m.
\end{split}
\end{equation}  
The profile of a mixture of Gamma functions has been studied in a detailed way in a recent paper \cite{GCC16}, 
to characterize the intensity of the bimodal profile. We report in Figure \ref{fg:bimodal} the resulting profiles for various choices of $\mu_S<\mu_R$, and $S^\infty$, $R^\infty$. Note that the mixture of the two inverse Gamma densities \fer{eq:fRS}  does not always result in an evident bimodal shape. Indeed, while the profile on the right of Figure \ref{fg:bimodal} is clearly bimodal, a different choice of parameters on the left produces a unimodal steady profile.

\section{Numerical simulations}\label{numerics}
In this section, we present some numerical examples to demonstrate the model's ability to describe different situations of wealth distribution in the presence of an epidemic phenomenon. We start from a validation of the Fokker-Planck limit by integrating the Boltzmann equations through a Monte Carlo method for the wealth distribution (see \cite{PR,PT13} for an introduction).
In the second example we show how the model is able to describe the increase in inequality in an economic system due to the advance of the epidemic. Among the observed effects we have a reduction (sometimes defined as "shrinking") of the so-called middle class. Finally, in the last example, we consider the case where social interactions depend on wealth and thus the spread of the epidemic impacts differently on different social classes. The disproportionate impact on the less wealthy social classes clearly emerges.
\subsection{Test 1: Asymptotic steady states}

\begin{figure}
	\begin{center}
		\includegraphics[scale=0.4]{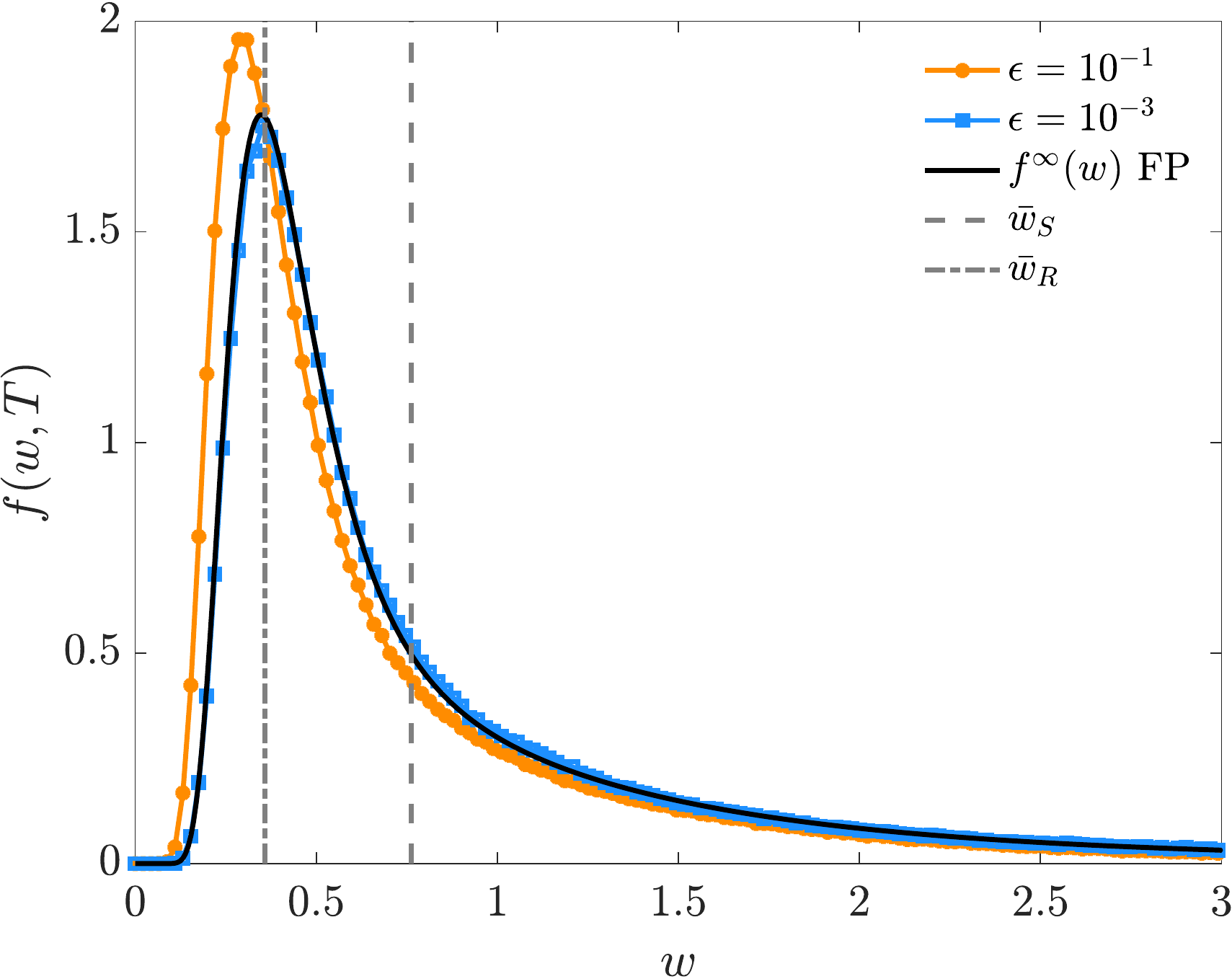}\hspace{0.2cm}
		\includegraphics[scale=0.4]{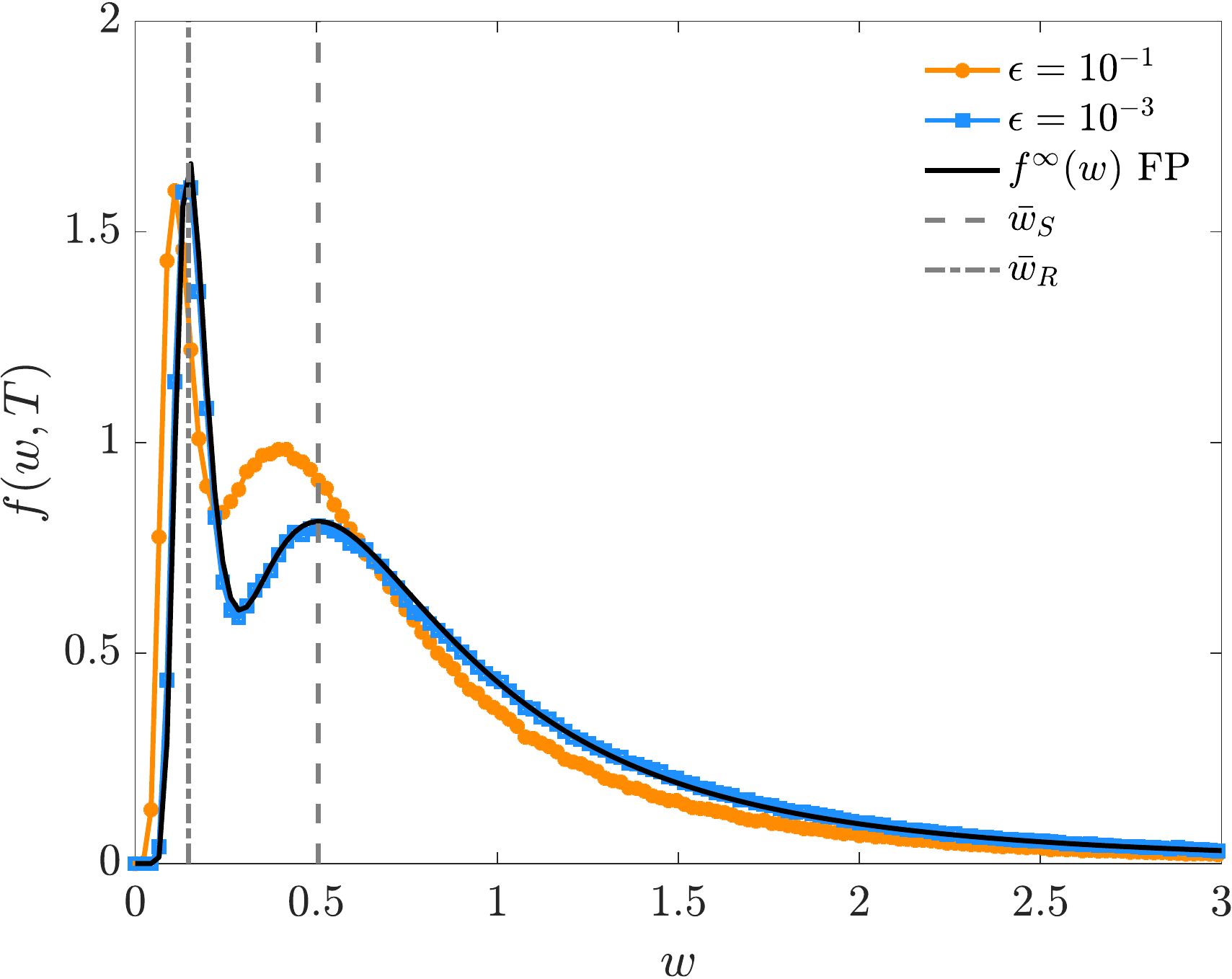}\\
	\end{center}
	\caption{\textbf{Test 1}. Comparison of the wealth distributions at the end of the epidemic for the kinetic system \eqref{eq:SIRw1}-\eqref{eq:SIRw3} with the explicit Fokker-Planck solution \eqref{eq:finf} with scaling parameters $\epsilon = 10^{-1}, 10^{-3}$. We considered $\mu_S = 2.5$ and $\mu_R = 7$ (left) or $\mu_R = 11$ (right). The asymptotic values of the epidemic dynamics are $S^\infty = 0.4$, $R^\infty = 0.6$ (left) and $S^\infty = 0.8$, $R^\infty = 0.2$ (right). }
	\label{fig:1}
\end{figure}

In this first test show that the asymptotic behaviour of the Boltzmann-type model  \eqref{eq:SIRw1}-\eqref{eq:SIRw3} under the quasi-invariant scaling  \eqref{eq:quasi}-\eqref{eq:quasi_epi} is consistently approximated for $\epsilon\ll 1$ by the stationary distribution of the system of Fokker-Planck equations \eqref{eq:SIRw1f}-\eqref{eq:SIRw3f}. The system of kinetic equations has been solved using  a direct simulation Monte Carlo approach with $N = 10^5$ particles.
% and we reconstruct the density on the wealth interval $[0,15]$. 

Given $f(w) = \dfrac{1}{2} \chi(w \in [0,2])$ a uniform distribution of wealth, we consider the following initial condition for the initial densities of the wealth of susceptible, infected and recovered 
\be\label{eq:test2_init}
\begin{split}
f_S(w,0) &= \rho_S f(w), \qquad f_I(w,0) =  \rho_I f(w),\qquad f_R(w,0) = \rho_R f(w),
\end{split}
\ee
where $\rho_I = 10^{-2}$, $\rho_R=0$ and $\rho_S = 1-\rho_I$ are such that $\rho_S+\rho_I + \rho_R = 1$. In order to compare our results with the analytic solutions of the Fokker-Planck model \ref{eq:fRS} we assume constant interaction rates $\beta=0.2$, $\gamma=0.1$ and a constant market standard deviation $\sigma = 0.1$. 
%As  discussed in Section \ref{sect:explicit}, in this case, we can compute explicit equilibrium densities of the Fokker-Planck model, i.e. $f^\infty(w) = f_S^\infty(w) + f_R^\infty(w)$ where $f^\infty_S(w)$ and $f^\infty_R(w)$ are defined in \eqref{eq:fRS}. 

In Figure \ref{fig:1}, we show the computed solution at time $T = 300$ of \eqref{eq:SIRw1}-\eqref{eq:SIRw3} in the scaling regimes $\epsilon = 10^{-1}, 10^{-3}$. The other computational parameters are consistent with the ones used to produce Figure \ref{fg:bimodal}. In the left image, we considered $\mu_S = 2.5$, $\mu_R = 7$ and an epidemic dynamics such that $S^\infty = 0.4$, $R^\infty = 0.6$, whereas the right plot considers a more amplified situation with $\mu_S = 2.5$ and $\mu_R = 11$ but a lighter epidemic spreading, which translates into $S^\infty = 0.8$ and $R^\infty = 0.2$. 

A direct comparison with the equilibrium density of the Fokker-Planck model confirms that if $\epsilon$ is small enough, Fokker-Planck's asymptotics provide a consistent approximation of the steady states of the Boltzmann dynamics. The emerging bimodal form of the asymptotic state is present in both cases represented, even if a greater discrepancy of the Pareto coefficients (right plot in Figure \ref{fig:1}) produces a more marked effect on the final distribution. To underline this we have drawn the maximum points of the distributions $f_S^\infty$, $f_R^\infty$ which are at $\bar w_S$, $\bar w_R$ defined in \eqref{eq:wbar_SR}, see dotted and dashed lines respectively.

\subsection{Test 2: Formation of wealth inequalities}

Next, we compare the evolution of the wealth distribution of the system under more realistic hypotheses on the dependence of the risk coefficient $\sigma$ from the epidemic spread as discussed in Section  \ref{sect:binary}. Hence, we consider the Boltzmann-type model \eqref{eq:SIRw1}-\eqref{eq:SIRw3} in the case of the following two infectious-dependent market risk coefficients
\begin{equation}
\label{eq:sigma_12}
\sigma_1(t)=\sigma_0 (1+\alpha I(t)), \qquad \sigma_2(t) = \sigma_0\left(1+\alpha \int_0^t I(\tau)d\tau\right),
\end{equation}
where $\alpha>0$, $\sigma_0>0$. In details, $\sigma_1(t)$ characterizes the instantaneous influence of the epidemic based on the observed number of infected, whereas $\sigma_2(t)$ takes into account possible long time memory effects on the market based on the epidemic impact.

As far as the epidemiological parameters are concerned, we consider a constant infection rate $\beta = $0.2 and a constant recovery rate $\gamma = $0.1. In the following, we will analyze the emerging inequalities assuming \eqref{eq:sigma_12} and two possible scenarios: the first considers a homogeneous population, $\lambda_S = \lambda_R$, which means that both populations share the same saving propensity, the second considers $\lambda_S>\lambda_R$ so that the saving propensity of the susceptible is lower than that of the recovered ones. 

\begin{figure}
	\begin{center}
	\subfigure[$\sigma_1$, $\lambda_S = \lambda_R$]{
		\includegraphics[scale=0.45]{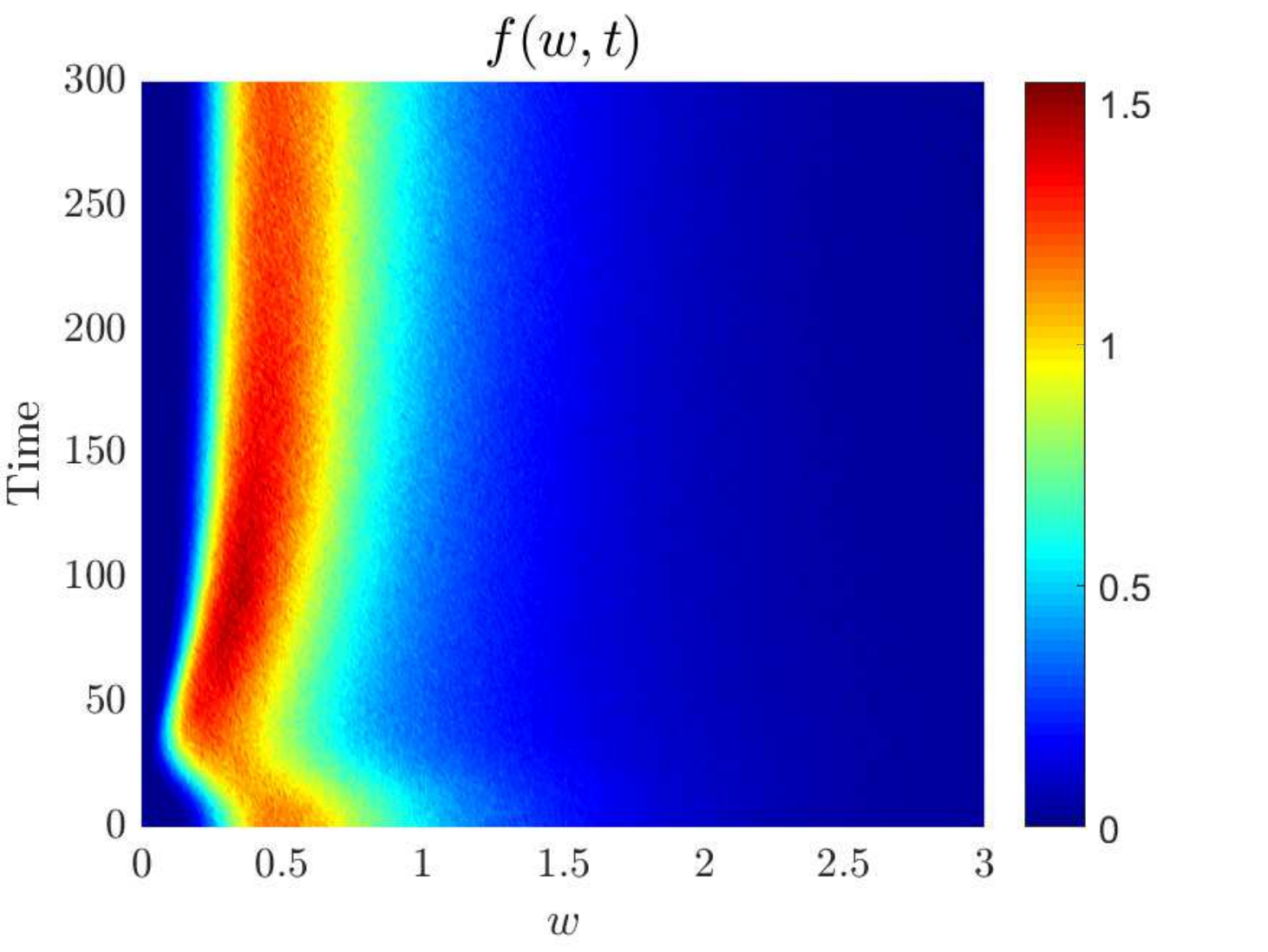}}
        	\subfigure[$\sigma_2$, $\lambda_S = \lambda_R$]{
		\includegraphics[scale=0.45]{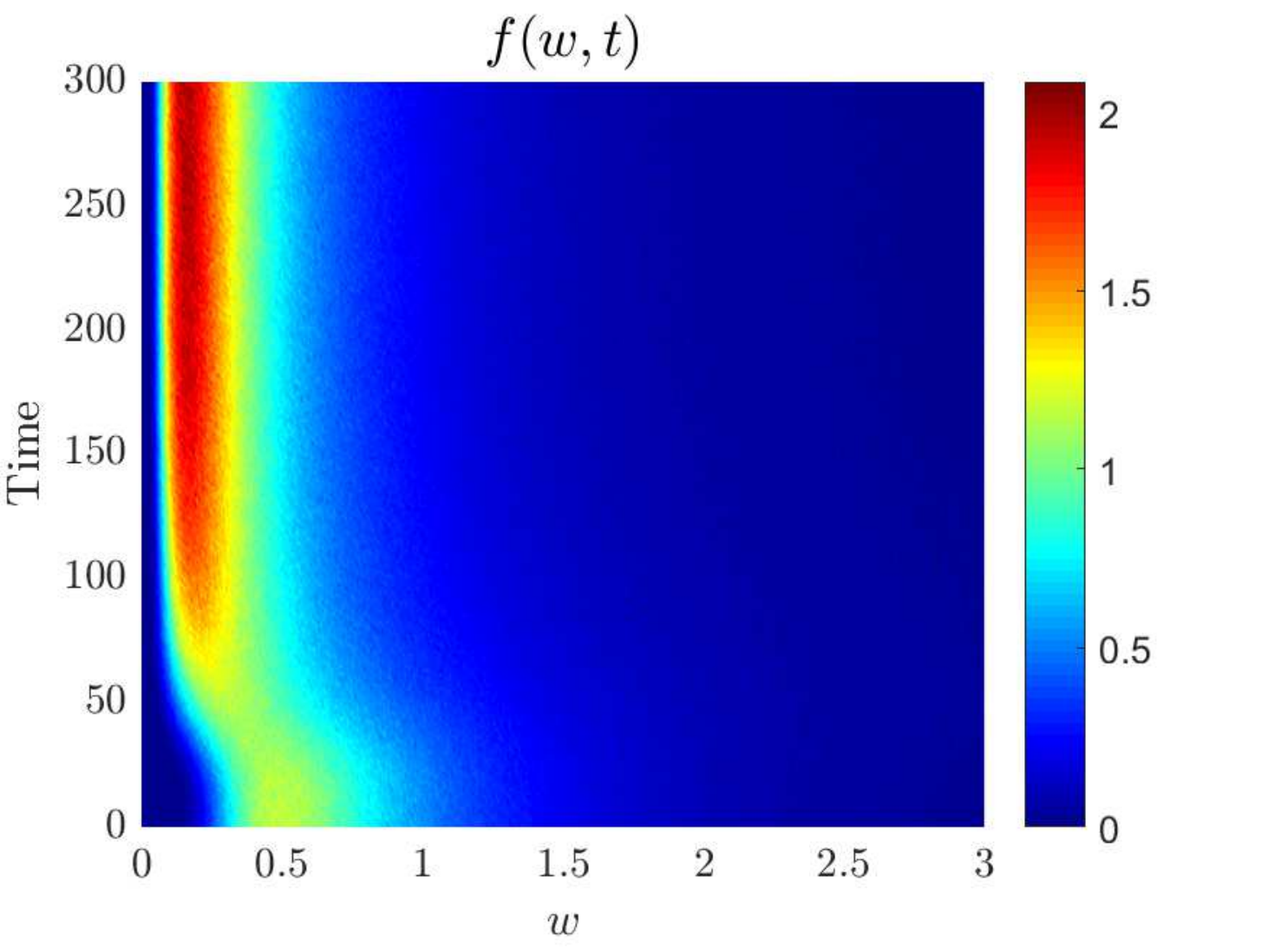}}\\
	\subfigure[$\sigma_1$, $\lambda_S<\lambda_R$]{
			\includegraphics[scale=0.45]{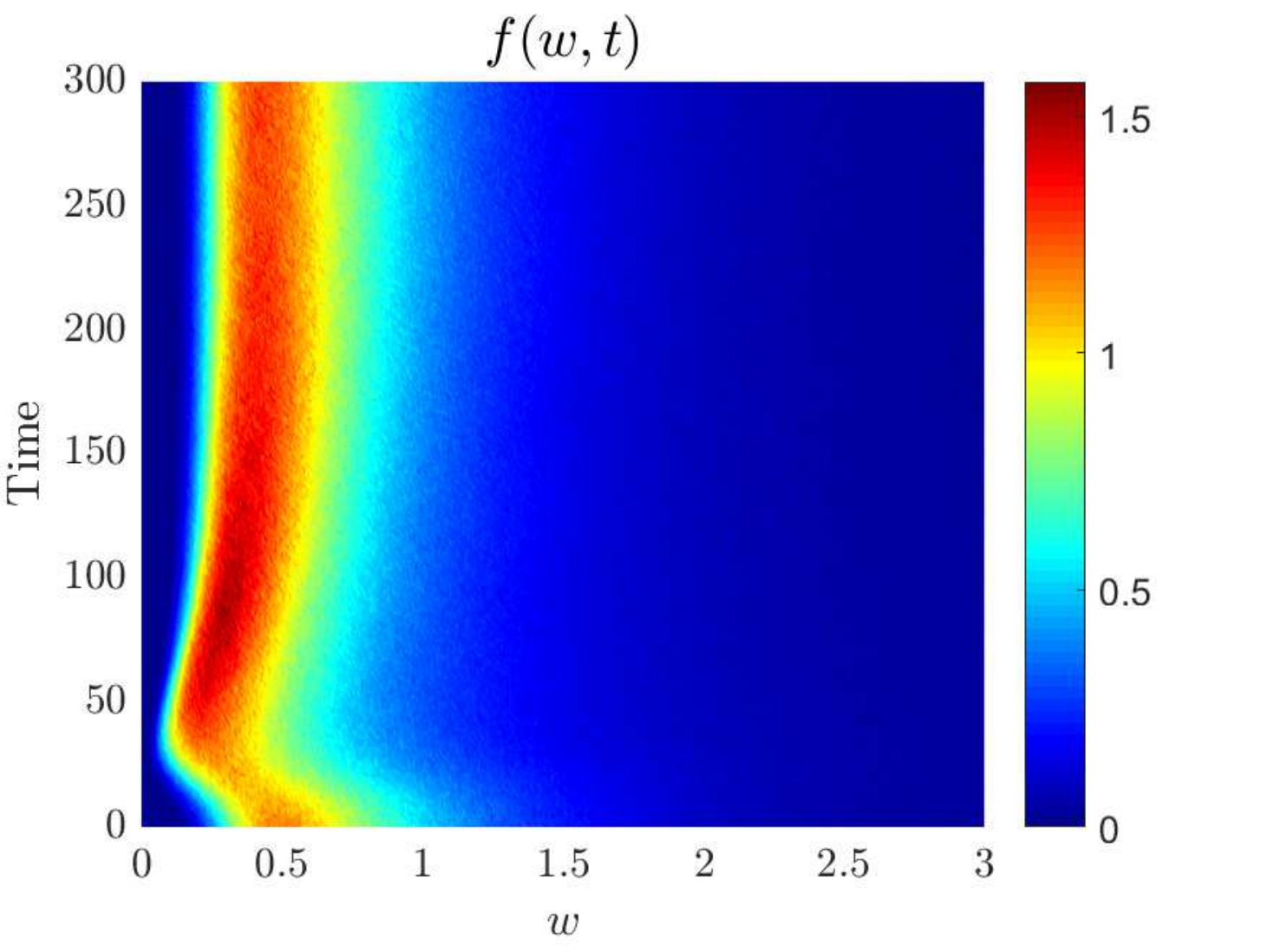}}
	\subfigure[$\sigma_2$, $\lambda_S<\lambda_R$]{
			\includegraphics[scale=0.45]{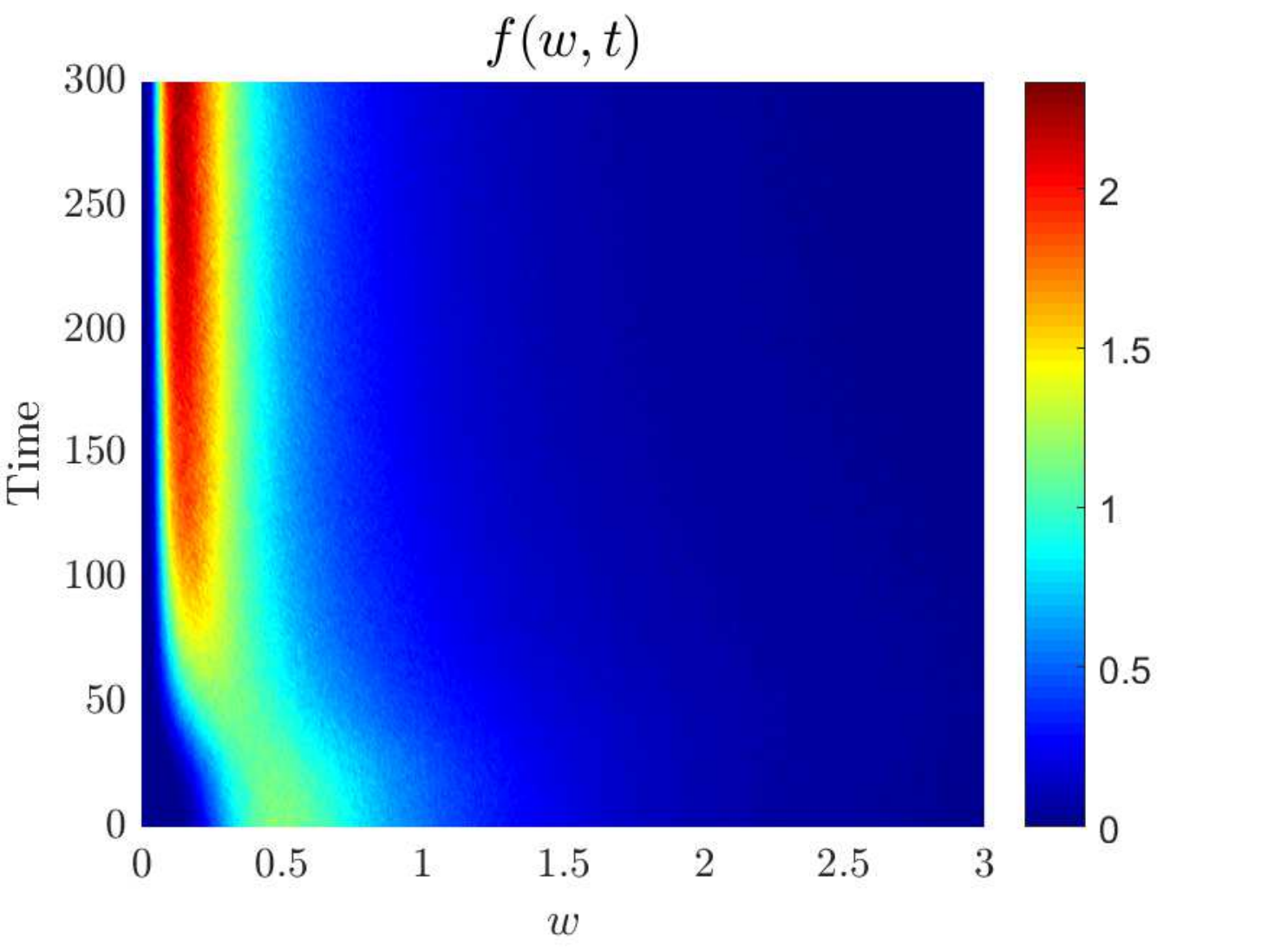}}
	\end{center}
	\caption{\textbf{Test 2}. Time evolution of the wealth distribution for the kinetic model \eqref{eq:SIRw1}-\eqref{eq:SIRw3} in the scaling $\epsilon = 5\times 10^{-3}$ with infectious-dependent market risk of the form \eqref{eq:sigma_12} with $\alpha =5$, $\sigma_0 = 0.1$. We considered $\lambda_S = \lambda_R = 0.1$ (top tow) and $\lambda_R = 2\lambda_S=0.1$ (bottom row).   }
	\label{fig:2}
\end{figure}

We consider, as initial distribution, an inverse Gamma distribution 
\be
f(w) = \dfrac{(\mu-1)^\mu}{\Gamma(\mu)} \dfrac{\textrm{exp} \left( -\frac{\mu-1}{w}\right)}{w^{1+\mu}}
\ee
with $\mu = 3$, representing an initial economic equilibrium state. Hence, we define, as before, the initial density of the wealth of susceptible, infected and recovered as in \eqref{eq:test2_init} where $\rho_I = 10^{-2}$, $\rho_R=0$ and $\rho_S = 1-\rho_I$ such that $\rho_S+\rho_I + \rho_R = 1$.  The total number of particles in the Monte Carlo method is $N = 10^5$.
In Figure \ref{fig:2} we represent the evolution of the total wealth distribution for both cases in \eqref{eq:sigma_12} starting from the initial distributions in \eqref{eq:test2_init}. In detail, the top line shows the case $\lambda_S = \lambda_R = 0.1$ and the bottom line $\lambda_R = 0.1 = 2\lambda_S$.

We can see that in both cases the pandemic can also strongly change the transitional wealth distribution regimes. In particular, in the case $\sigma_1(t)$ the solution of the kinetic model shows a non-monotonous dynamics such that, if $\lambda_R = \lambda_S$, the equilibrium distribution coincides with the initial state $f(w,0) = f_S(w,0) + f_R(w,0) + f_I(w,0)$, see \eqref{eq:test2_init}. On the contrary, if $\lambda_R>\lambda_S$, we expect a different final state to occur as shown in the right images of Figure \ref{fig:2}.  In particular, in the case $\sigma_2(t)$, that is when the epidemic memory is present, the dynamics both for $\lambda_R = \lambda_S$ and for $\lambda_R> \lambda_S$ highlights the trend towards an equilibrium centered on lower wealth values.

\begin{figure}
\centering
\includegraphics[scale = 0.5]{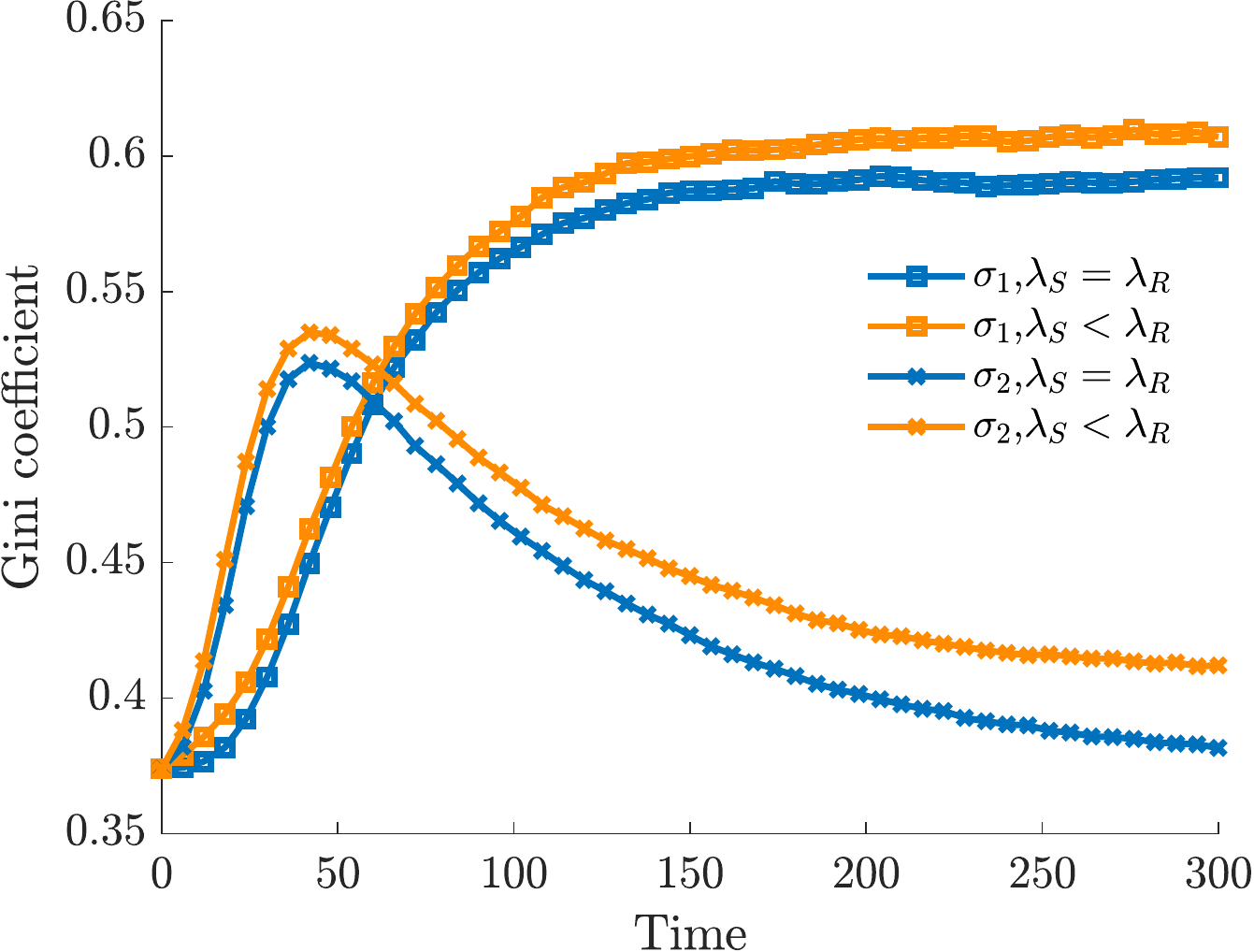}
\includegraphics[scale = 0.5]{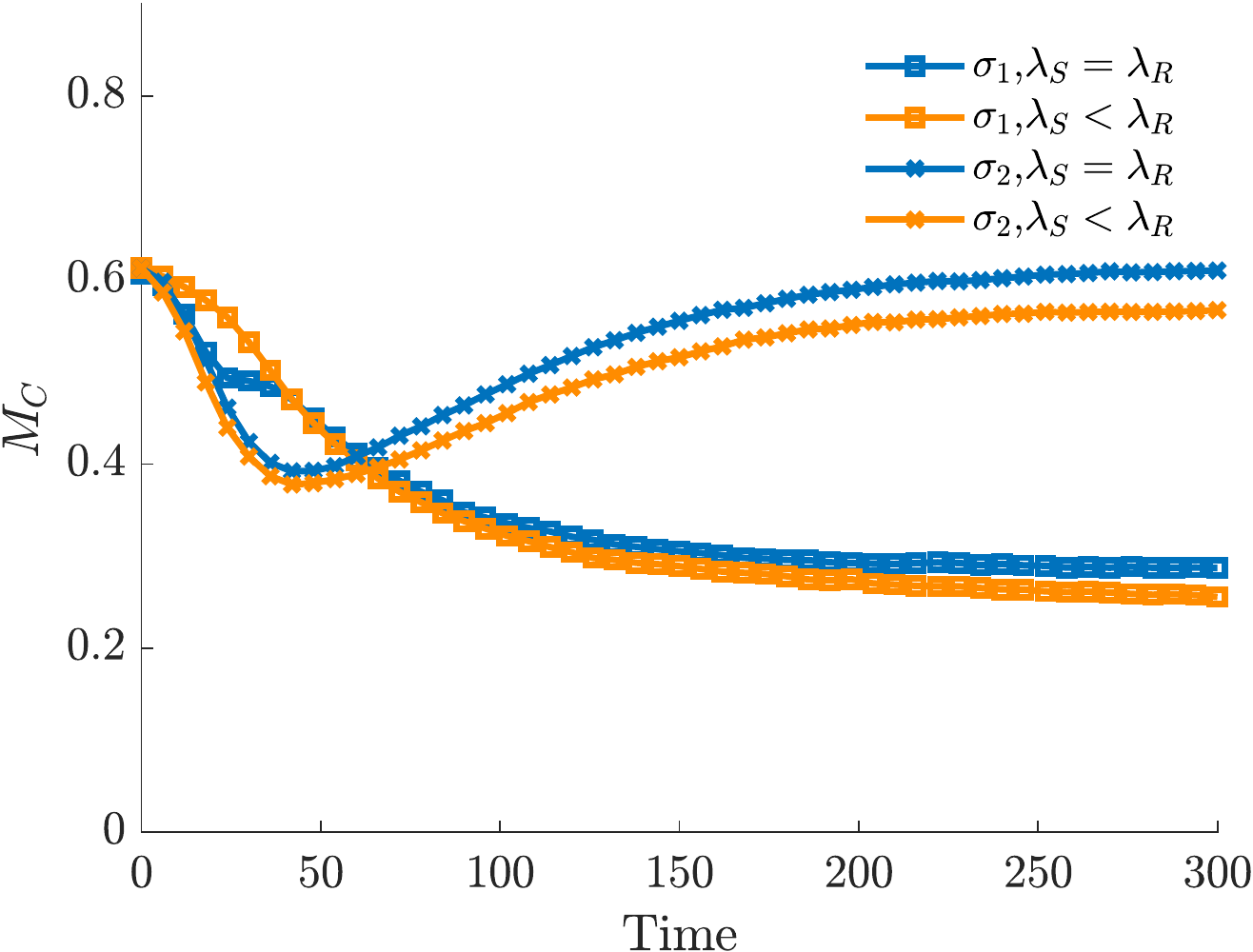}
\caption{\textbf{Test 2}. Behavior of the Gini index (left) and of the middle class fraction (right) defined in \eqref{eq:MC} during the outbreak of the epidemic for the different risk measures in \eqref{eq:sigma_12} with $\alpha =5$, $\sigma_0 = 0.1$. }
\label{fig:3}
\end{figure}

To get a more detailed view of the emerging equilibria, we resort to the Gini index calculation, see \cite{Gini, DPT}. This value should be understood as a measure of a country's wealth inequality and varies in $[0,1]$, where $0$ indicated perfect equality and $1$ perfect inequality. In modern economies, a Gini index between $[0.2,0.5]$ is often observed. In Figure \ref{fig:3} we represent the evolution of the Gini index $G_1$ for the tests considered in Figure \ref{fig:2}. We clearly observe an inequality of wealth that grows with the epidemiological dynamics. Moreover, even in the case of $\sigma_1$ with $\lambda_S = \lambda_R$, where these effects are absorbed in the long-lasting trends, the recovery of the economy occurs at a much lower rate than the worsening rate.

Epidemiological dynamics may translate into additional wealth inequalities, in particular we can measure the evolution of the total number of individuals belonging to the middle class. Although there are several ways to give a technical definition of the middle class, it is often more of an idea or estimate than a fixed number. Generally speaking, the middle class is loosely defined as those who fall into the middle group of workers compared to the bottom $20\%$ or top $20\%$. We can define it using an interval $[w_L,w_R]$ such that
\be\label{eq:classes} 
\int_{0}^{w_L} f(w,0) \approx 0.2,\qquad \int_{w_R}^{\infty} f(w,0) \approx 0.2,
\ee 
and computing the time evolution of
\be\label{eq:MC}
M_C(t) = \int_{w_L}^{w_R} f(w,t)\,dw,  
\ee
gives us an estimate of the percentage of people living in middle-income households. In Figure \ref{fig:3} (right plot), we represent the evolution of $M_C(t)$ corresponding to the considered $\sigma_1(t)$, $\sigma_2(t)$. We can clearly see how the emerging inequalities mainly affect the middle class, which is constantly decreasing in the case of $\sigma_2$ and undergoes a transitory decrease for $\sigma_1$. In particular, in this last scenario and in the $\lambda_S<\lambda_R$ regime, at the end of the epidemic dynamics only a partial recovery to the original pre-epidemic level is observed.

%\subsection{Test 3: Importance of a state health care system}
%Consider the recovery rate as a function of the wealth, therefore taking into account the possible lack of a health care system. For example, having insurance is essential in the American health care system. Those who aren’t insured can accrue overwhelming medical bills or face outright refusal when they do seek care. As a result, this increases the number of active infected people and thus reduces their recovery rate.
%This is modeled using a sigmoid function $S(\cdot)\in (0,1)$ in the form
%\begin{equation}
%\gamma(w)={\gamma_0}{S(w)},
%\label{eq:gammaI}
%\end{equation}
%for example $S(w)=w/(1+w)$. Show that we have an increase of infected people in the poorer classes of the population. Use just one given value of $\varepsilon$. Plot the wealth distribution of infected people for $\gamma=\gamma_0$ and $\gamma$ defined from (\ref{eq:gammaI}) to see the effect in time.

\subsection{Test 3: Impact of epidemic on wealth status}
\begin{figure}
	\centering
	\includegraphics[scale = 0.5]{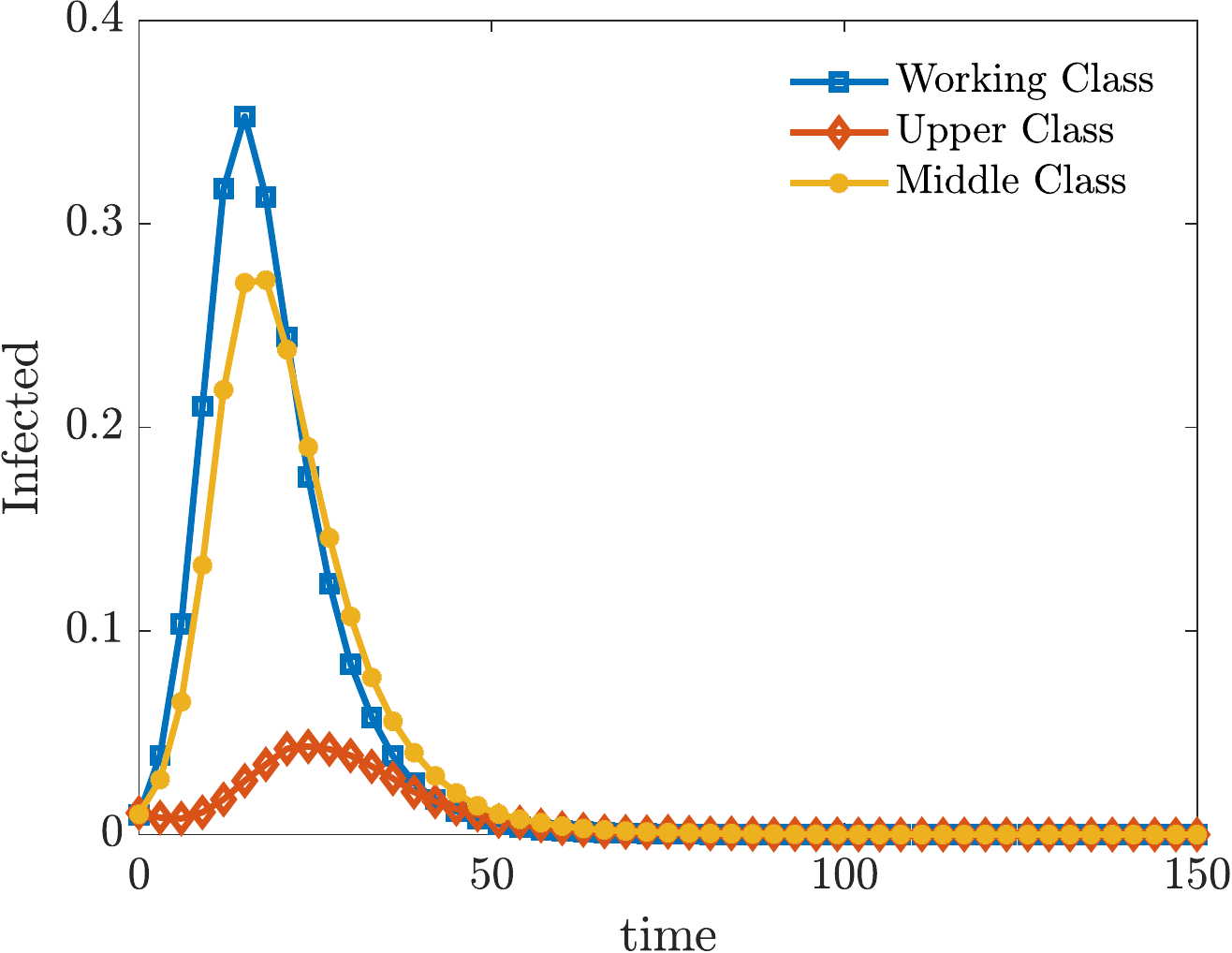}
	\includegraphics[scale = 0.5]{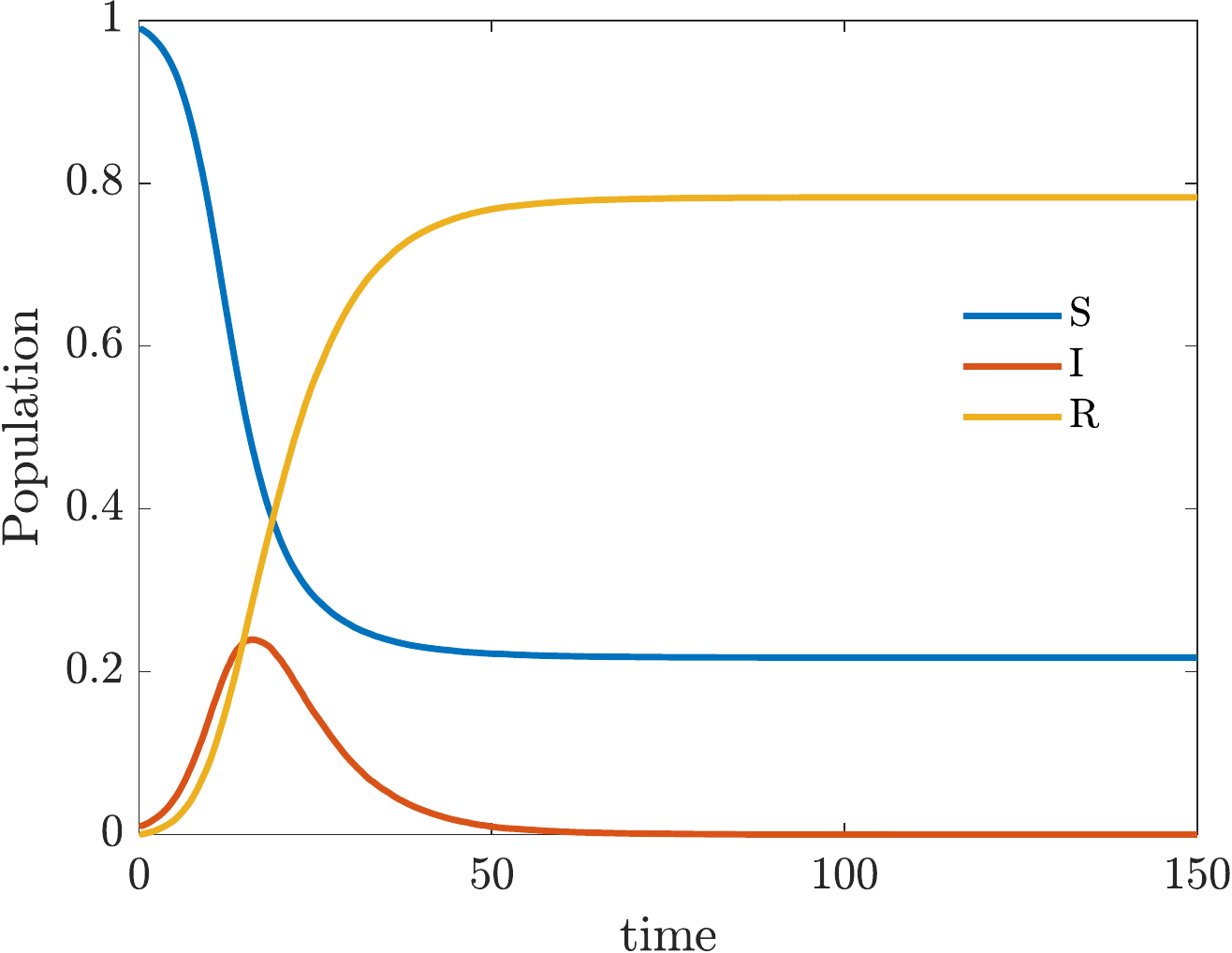}
	\caption{\textbf{Test 3}. Evolution of the fraction of infectious individuals belonging to different social classes (left) and behavior of the corresponding densities for susceptibles, infectious and recovered (right) in time.}
	\label{fig:4}
\end{figure}

\begin{figure}
	\centering
	\includegraphics[scale = 0.5]{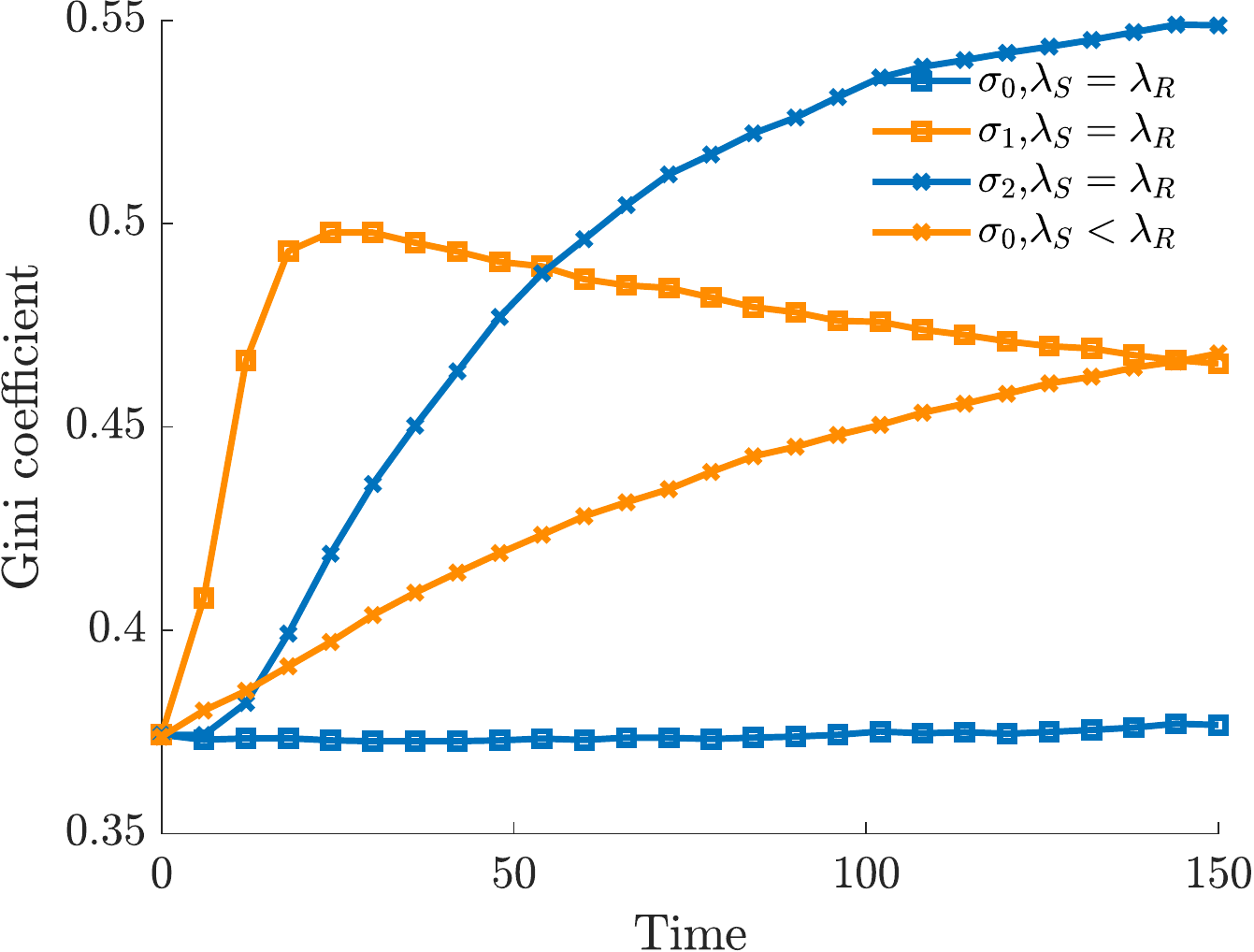}
	\includegraphics[scale = 0.5]{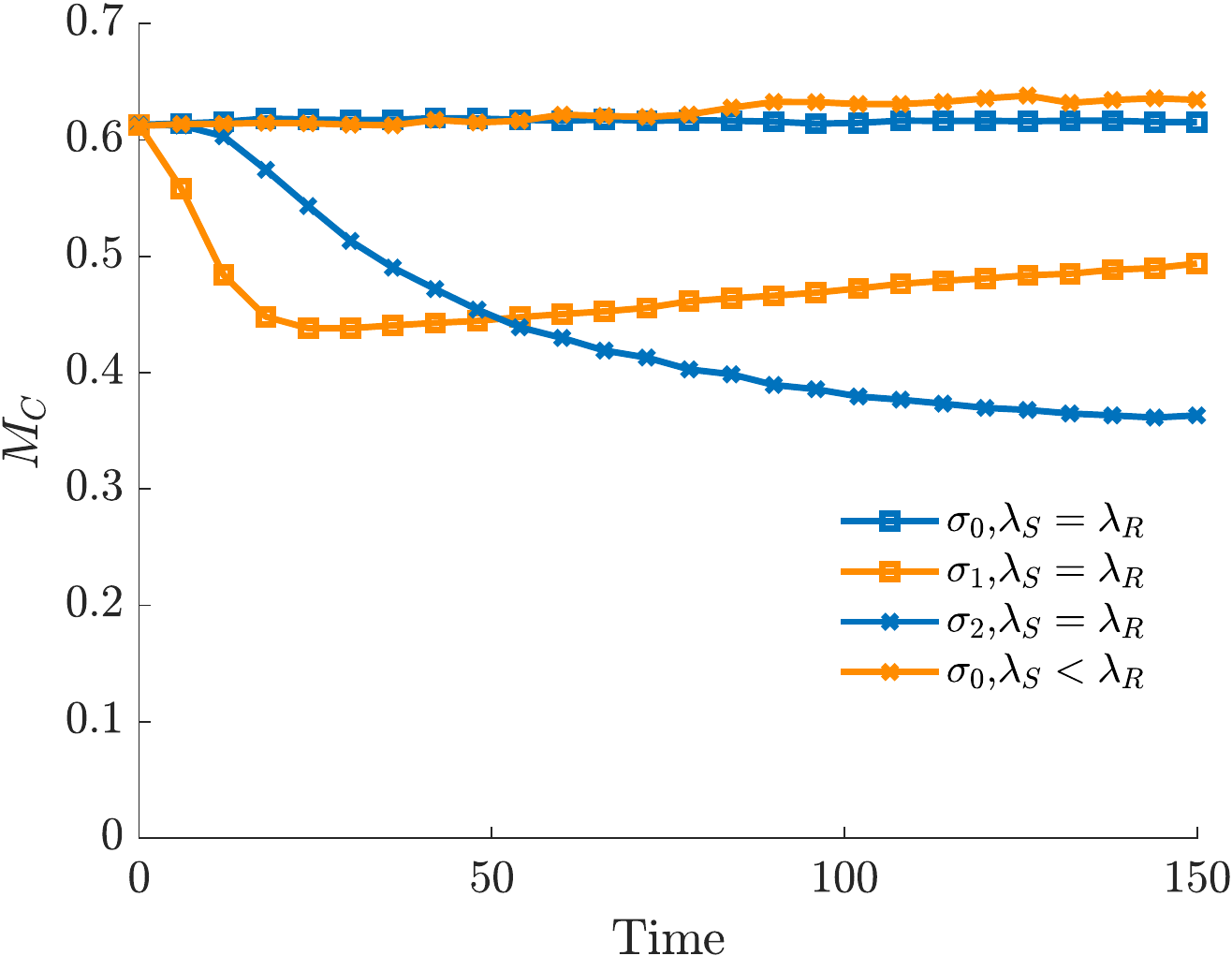}
	\caption{\textbf{Test 3}. Behavior of the Gini index (left) and of the middle class fraction (right) defined in \eqref{eq:MC} during the outbreak of the epidemic for  different risk measures and values of $\lambda_S$ and $\lambda_R$.}
	\label{fig:5}
\end{figure}

%As the US climbed to more than $20,000$ coronavirus deaths state health officials grappled with its disproportionate impact on black Americans. African Americans face a higher risk of exposure to the virus, mostly on account of concentrating in urban areas and working in essential industries. Only $20\%$ of black workers reported being eligible to work from home, compared with about $30\%$ of their white counterparts, according to the Economic Policy Institute.
In the last example, we consider an interaction term dependent from the wealth of the agents. The interaction function takes into account the fact that interactions occur more frequently between people of the same social status and are higher for people belonging to the working class 
\begin{equation}
\beta(w,w_*)={b(w)b(w_*)}\Psi(|w-w_*| \leq \Delta),
\end{equation}
where $b(w) = {\beta_0}/{(1+w^\alpha)}$, and $\Psi(\cdot)$ is the indicator function. In the following we assume $\alpha=\Delta=2$. In Figure \ref{fig:4}, we show the course of the infection over time. The image on the right shows the relative number of sensitive, recovered and infected subjects, while the image on the left shows the number of infected by social class. These are divided into three classes: working class, middle class and upper class as defined by \eqref{eq:classes} and \eqref{eq:MC}. The recovery rate is set equal for the whole population to $ \gamma=0.3$. We see clearly how the infection affects people with a lower level of wealth. As for wealth, we choose constant transaction coefficients $\lambda_S=\lambda_R$ as well as constant market risk $\sigma=0.1$. In Figure \ref{fig:5}, we report the Gini index and the trend of the percentage of agents belonging to the middle class for four different scenarios. These are in addition to the one described above regarding the influence of epidemics on market risk as in \eqref{eq:sigma_12}. First we consider the last case when the agents behave differently based on whether they belong to sensitive, infected or recovered subjects with $\lambda_S=0.025<\lambda_R=0.1$, and constant $\sigma=0.1$. In this scenario scenario, the disease does not change Gini's index. On the other hand, for the other three cases the index shows a clear impoverishment of the population. In particular, for the case in which the market is affected by the instantaneous spread of the epidemic in absence of memory effect, the index trend shows that after some time the level of wealth returns to the original situation. This is not the case when the market maintains the memory of the epidemic and when agents behave differently according to their state of health. Looking at the number of agents belonging to the middle class, we realize that the number of people who can be considered as belonging to this class decreases with time in only two situations, while it seems to remain lower than before the epidemic only if the market keeps the memory of the spread of the disease.  

\section{Concluding remarks}
In this work we have introduced a first example of an economic model in the presence of an epidemic phenomenon. The recent COVID-19 pandemic has in fact highlighted the importance of having models capable of describing the social and economic impact of the spread of the disease, albeit in a simplified way. The increase in social inequalities, the reduction of the middle class and the diversified impact of the epidemic on different social classes are phenomena typically observed in these conditions and that we have shown the model is able to describe. 
In addition to the modelling aspects, in the present work we included a qualitative analysis which shows that in a certain range of the parameters the kinetic wealth distribution system is well-posed, it has a unique solution for a large class of initial data, and  the large-time behavior of this solution is found in a universal steady state.  Explicitly computable stationary solutions have been evaluated by resorting to an asymptotic description of the kinetic system, which is expressed by a system of  Fokker-Planck type equations. 
Finally, we note that the modelling approach just described can be easily adapted to more realistic compartmental models in epidemiology such as SEIR and/or MSEIR, including additional social effects, such as the age of agents \cite{HWH00}. Nevertheless, the fundamental aspects of the interaction between wealth and spread of infectious disease will not change. These generalizations and related numerical simulations will be the subject of future research.

%%%%%%%%%%%%%%%%%%%%%%%%%%%%%%%%%%%%%%%%%%%%%%
\appendix
\section{Proof of Theorem \ref{thm:main}} \label{sec:appendix}

%We present in this Appendix the details of the proof of Theorem \ref{thm:main}. 
We consider the system in Fourier version \fer{eq:SIRf1}-\fer{eq:SIRf3} of Section \ref{sec:analysis}. 
Owing to mass conservation, for $H\in  \{S,I,R\}$  we can write in a compact form
\be\label{eq:semplice}
\sum_{J\in \{S,I,R\}} \hat Q(\fH,\fJ)(\xi,t) =   \hat Q_+(\fH)(\xi,t) - \fH(\xi,t).
\ee
In \fer{eq:semplice},  the gain operators $\hat Q_+$,  for $H\in  \{S,I,R\}$, are defined by 
\be\label{eq:q+}
\hat Q_+(\fH)(\xi,t) = \sum_{J\in \{S,I,R\}} \langle\fH (A_{HJ}\xi, t)\rangle \fJ(\lambda_J\xi,t),
\ee
where $A_{HJ}$ has been defined in \eqref{eq:breve}. Using notation \fer{eq:q+} into  system \fer{eq:SIRf1}-\fer{eq:SIRf3} allows to write it in the form
\begin{eqnarray}
\frac{\partial \fS(\xi,t)}{\partial t}+ \fS(\xi,t) &=& -\beta I(t) \fS(\xi,t) + \hat Q_+(\fS)(\xi,t)
\label{eq:f1}\\
\frac{\partial \fI(\xi,t)}{\partial t} + \fI(\xi,t)&=& \beta I(t) \fS(\xi,t) - \gamma \fI(\xi,t) +\hat Q_+(\fI)(\xi,t)
\label{eq:f2}\\
\frac{\partial \fR(\xi,t)}{\partial t}+ \fR(\xi,t) &=& \gamma \fI(\xi,t) + \hat Q_+(\fR)(\xi,t)
\label{eq:f3}
\end{eqnarray}
Notice that not all the coefficients of the terms on the right-hand sides are positive. However, since  $I(t) \le 1 $, and $\beta,\gamma <1$, we can obtain positivity of coefficients by adding $ \fS(\xi,t)$ to \fer{eq:f1}, and similarly $ \fI(\xi,t)$ to \fer{eq:f2} and $ \fR(\xi,t)$ to \fer{eq:f3}. This allows to obtain the equivalent system
\begin{eqnarray}
\frac{\partial \fS(\xi,t)}{\partial t}+ 2\fS(\xi,t) &=& (1-\beta I(t)) \fS(\xi,t) + \hat Q_+(\fS)(\xi,t)
\label{eq:g1}\\
\frac{\partial \fI(\xi,t)}{\partial t} + 2\fI(\xi,t)&=& \beta I(t) \fS(\xi,t) +(1- \gamma) \fI(\xi,t) +\hat Q_+(\fI)(\xi,t)
\label{eq:g2}\\
\frac{\partial \fR(\xi,t)}{\partial t}+ 2\fR(\xi,t) &=& \fR(\xi,t) +\gamma \fI(\xi,t) + \hat Q_+(\fR)(\xi,t)
\label{eq:g3}
\end{eqnarray}
in which all coefficients on the right-hand sides are positive.

Let $\fJ(t)$ and $\gJ(t)$, $t >0$  denote two solutions to the SIR system corresponding to initial data with the same masses $J(t=0)$ and moments $m_J(t=0)$, $J \in \{S,I,R\}$.
Then, the evolution of masses and mean values, as given by \fer{eq:SIR1} - \fer{eq:SIR3} (respectively \fer{eq:SIRm1} - \fer{eq:SIRm3}) guarantee that both $f_J(t)$ and $g_J(t)$, $J \in \{S,I,R\}$,  have the same masses and mean values at any subsequent time $t>0$.  

Our objective is to investigate the time behavior of the $d_2$-metric, as defined in \fer{eq:ds}. Let us fix $s=2$, and let us define for $J \in \{S,I,R\}$
\be\label{eq;ds}
h_J(\xi,t)  =  \frac{\fJ(\xi) -\gJ(\xi)}{|\xi|^2}.
\ee
The functions $\hJ$ satisfy the system
\begin{eqnarray}
\frac{\partial \hS(\xi,t)}{\partial t} + 2\hS(\xi,t)  &=& (1-\beta I(t)) \hS(\xi,t) + L_+(\fS)(\xi,t)
\label{eq:hh1}\\
\frac{\partial \hI(\xi,t)}{\partial t} + 2\hI(\xi,t)&=& \beta I(t) \hS(\xi,t) +(1 - \gamma) \hI(\xi,t) + L_+(\fI)(\xi,t)
\label{eq:hh2}\\
\frac{\partial \hR(\xi,t)}{\partial t} + 2\hR(\xi,t)&=& + \hR(\xi,t) + \gamma \hI(\xi,t) + L_+(\fR)(\xi,t).
\label{eq:hh3}
\end{eqnarray}
In system \fer{eq:hh1}-\fer{eq:hh3} the functions $L_+(\fH)(\xi,t)$, with $H \in \{S,I,R\}$ are expresses by
\be\label{eq:L+}
L_+(\fH)(\xi,t) = \frac{\hat Q_+(\fH)(\xi,t) -\hat Q_+(\gH)(\xi,t)}{|\xi|^2}.
\ee
Owing to the definition of $\hat Q_+$ we argue that $L_+(\fH)$ is the sum of three similar terms, each one given by a product of two Fourier transforms. For $H,J \in \{S,I,R\}$ these terms read
\[
\frac{\big\langle \fH (A_{HJ}\xi, t) \fJ(\lambda_J\xi,t)-  \gH (A_{HJ}\xi, t)\gJ(\lambda_J\xi,t)\big\rangle}{|\xi|^2},
\]
where for convenience we put outside the mean value $\langle\cdot\rangle$.
For products of this type we can handle the difference in a suitable way  \cite{DMT, MT08, PT13} to obtain
\begin{align*}
& \bigg |  \frac{\big\langle \fH (A_{HJ}\xi, t) \fJ(\lambda_J\xi,t) -\gH (A_{HJ}\xi, t)\gJ(\lambda_J\xi,t)\big\rangle}{|\xi|^2} \bigg |
\\
&\quad\leq \bigg\langle |\fH (A_{HJ}\xi, t)| \bigg | \frac{ \fJ(\lambda_J\xi,t) -\gJ(\lambda_J\xi,t)}{|\lambda_J \xi|^2} \bigg |\lambda_J^2 \bigg\rangle  \\
&\qquad\qquad+ \bigg\langle |\gJ (\lambda_{J}\xi, t)| \bigg | \frac{ \fH(A_{HJ}\xi,t) -\gH(A_{HJ}\xi,t)}{|A_{HJ}\xi|^2} \bigg |A_{HJ}^2 \bigg\rangle \\
&\quad\leq H(t) \sup_\xi  \bigg | \frac{ \fJ(\xi,t) -\gJ(\xi,t)}{|\xi|^2} \bigg|\lambda_J^2 + J(t) \sup_\xi  \bigg | \frac{ \fH(\xi,t) -\gH(\xi,t)}{|\xi|^2} \bigg|\langle A_{HJ}^2\rangle \\
&\quad = \lambda_J^2 H(t) \|h_J(t)\|_\infty + \langle A_{HJ}^2 \rangle J(t) \|h_H(t)\|_\infty.
\end{align*}
Hence, for  $H \in \{S,I,R\}$ we conclude with the bound
\be\label{eq:bL+}
\|L_+(\fH)(t)\|_\infty  \le H(t)\sum_{J\in \{S,I,R\}} \lambda_J^2  \|h_J(t)\|_\infty + \|h_H(t)\|_\infty \sum_{J\in \{S,I,R\}} \langle A_{HJ}^2 \rangle J(t). 
\ee
Let us multiply both sides of \fer{eq:hh1} - \fer{eq:hh3} by $e^{2t}$. Since the coefficients on the right-hand sides are positive, we obtain
\begin{eqnarray*}
	\frac{\partial [\hS(\xi,t)e^{2t}] }{\partial t} &\le& (1-\beta I(t)) \|\hS(t)e^{2t}\|_\infty  +  \|L_+(\fS)(t)e^{2t}\|_\infty 
	\label{eq:hm1}\\
	\frac{\partial [\hI(\xi,t)e^{2t}] }{\partial t}&\le& \beta I(t) \|\hS(t) e^{2t}\|_\infty +(1 - \gamma) \|\hI(t) e^{2t}\|_\infty + \|L_+(\fI)(t)e^{2t}\|_\infty 
	\label{eq:hm2}\\
	\frac{\partial [\hR(\xi,t)e^{2t}] }{\partial t}&\le& + \|\hR(t)e^{2t}\|_\infty + \gamma \|\hS(t) e^{2t}\|_\infty +  \|L_+(\fR)(t)e^{2t}\|_\infty 
	\label{eq:hm3}
\end{eqnarray*}
Let us integrate both sides of the equations of the system from $0$ to $t$. By taking the supremum on both sides we finally get 
\begin{eqnarray*}
	\|\hS(t)e^{2t}\|_\infty &\le&  \|\hS(0)\|_\infty + \int_0^t \left[ (1-\beta I(s)) \|\hS(s)e^{2s}\|_\infty +  \|L_+(\fS)(s)e^{2s}\|_\infty  \right] \, ds
	\\
	\|\hI(t)e^{2t}\|_\infty &\le&  \|\hI(0)\|_\infty + \int_0^t \left[ \beta I(s) \|\hS(s) e^{2s} \|_\infty +(1 - \gamma) \|\hI(s) e^{2s}\|_\infty +\|L_+(\fI)(s)e^{2s}\|_\infty \right]\, ds
	\\
	\|\hR(t)e^{2t}\|_\infty &\le&  \|\hR(0)\|_\infty + \int_0^t \left[ \|\hR(s)e^{2s}\|_\infty(s) + \gamma \|\hS(s) e^{2s}\|_\infty + \|L_+(\fR)(s)e^{2s}\|_\infty\right]\, ds.
\end{eqnarray*}
Now we define 
\begin{equation*}\label{eq:D+}
D(t) = \sum_{J\in \{S,I,R\}} \|\hJ(t)e^{2t}\|_\infty. 
\end{equation*}
Hence, summing up  the three equations in the system we obtain
\be\label{eq:boundD}
D(t) \le D(0) + \int_0^t \big[ D(s) + \sum_{H\in \{S,I,R\}}\|L_+(\fH)(s)e^{2s}\|_\infty \big]\, ds
\ee
It is now clear that  the large-time behavior of $D(t)$ in \fer{eq:boundD} depends heavily on the characteristics of the wealth operators, here represented in the sum of the  $\|L_+(\fH)(s)e^{2s}\|_\infty$. 
Making use of  \fer{eq:bL+} and of mass conservation,  for $t >0$ we have
\[
\begin{split}
&\sum_{H\in \{S,I,R\}}\|L_+(\fH)(t)\|_\infty  \\
&\qquad =  \sum_{H,J\in \{S,I,R\}} \lambda_J^2 H(t) \|h_J(t)\|_\infty + \sum_{H,J\in \{S,I,R\}} \langle A_{HJ}^2 \rangle J(t) \|h_H(t)\|_\infty \\
& \qquad = \sum_{H,J\in \{S,I,R\}} \big[ \lambda_J^2 + \langle A_{JH}^2 \rangle \big] H(t) \|h_J(t)\|_\infty\\
&\qquad \le \max_ {H,J\in \{S,I,R\}}\big[ \lambda_J^2 + \langle A_{JH}^2 \rangle \big] \sum_{J\in \{S,I,R\}}  \|h_J(t)\|_\infty. 
\end{split}
\]
Hence, if condition \fer{eq:regola} holds,  
\be\label{eq:finale}
\sum_{H\in \{S,I,R\}}\|L_+(\fH)(t)e^{2t}\|_\infty \le \nu D(t),
\ee
with $\nu <1$. At this point, Gronwall inequality applied to \fer{eq:boundD} implies
\[
D(t) \le D(0) \exp\{ (1+\nu)t\} ,
\]
and, consequently
\be\label{eq:decay} 
\sum_{J\in \{S,I,R\}} \|\hJ(t)\|_\infty \le  \sum_{J\in \{S,I,R\}} \|\hJ(0)\|_\infty \exp\{ -(1-\nu)t\}.
\ee
Thus, since for $J\in \{S,I,R\}$ 
\[
\|\hJ(t)\|_\infty = d_2(f_J(t), g_J(t))
\]
the $d_2$-metrics \fer{eq:ds} between the two solutions $f_J(v,t)$ and $g_J(v,t)$, $J\in \{S,I,R\}$, which at time $t=0$ have the same mass and the same mean value, decay to zero exponentially at a rate $1-\nu$. This concludes the proof of Theorem \ref{thm:main}.

%%%%%%%%%%%%%%%%%%%%%

%%%%%%%%%%%%%%%%%%%%%

\section*{Acknowledgement} This work has been written within the
activities of GNFM and GNCS groups of INdAM (National Institute of
High Mathematics), and partially supported by Italian Ministry of Education, University and Research (MIUR) - PRIN project 2017 ``Optimal mass
transportation, geometrical and functional inequalities with applications''.
The research of G. Toscani and M. Zanella was partially supported by
MIUR - Dipartimenti
di Eccellenza Program (2018--2022) - Dept. of Mathematics ``F.
Casorati'', University of Pavia. G. Dimarco and L. Pareschi acknowledge the partial support of  MIUR - PRIN Project 2017 ``Innovative numerical methods for evolutionary partial differential equations and applications''.

%%%%%%%%%%%%%%%%%%%%%%%%%%%%%%%%%%%%%%%%


\begin{thebibliography}{99.}

\bibitem{Ang} J. Angle, {The surplus theory of social stratification and the size distribution of personal wealth}, \emph{Social Forces} \textbf{65}:
 293--326,  1986.
 
\bibitem{Ang1} J. Angle, {The inequality process as a wealth maximizing process}., \emph{Physica A} \textbf{367}:  
388--414, 2006.

\bibitem{BG} 
L.Baringhaus, and R. Gr\"ubel, On a class of characterization problems for random convex combinations. \emph{Ann. Inst. Statist. Math} \textbf{49} 555--567, 1997.

\bibitem{BaTo}
F.~Bassetti, and G.~Toscani,
\newblock {Explicit equilibria in a kinetic model of gambling},
\newblock \emph{ Phys. Rev. E }, {\bf 81}: 066115, 2010.

\bibitem{BaTo2}
F.~Bassetti, and G.~Toscani,
\newblock {Explicit equilibria in bilinear kinetic models for socio-economic
  interactions},
\newblock \emph{ESAIM: Proc. and Surveys},  {\bf 47}: 1--16, 2014.

\bibitem{BST} 
M. Bisi, G. Spiga, and G. Toscani,  {Kinetic models of conservative economies with wealth redistribution},
\emph{Commun. Math. Sci.}, \textbf{7}: 901--916, 2009. 

\bibitem{BM}
J.F. Bouchaud,  and  M. M\'ezard, {Wealth condensation in a
simple model of economy}, \emph{Physica A\/}, {\bf 282}: 536--545, 2000.

\bibitem{CS78} V. Capasso and G. Serio. A generalization of the Kermack-McKendrick deterministic epidemic model. {\em Math. Biosci.} 42--43, 1978.

\bibitem{Gini}
S. R. Chakravarty, {Ethical Social Index Numbers},
\newblock (Springer-Verlag 1990).

\bibitem{Ch02}
A.~Chakraborti, {Distributions of money in models of market economy}, \emph{ Int. J. Modern Phys. C}, \textbf{13}: 
1315--1321, 2002.

\bibitem{ChaCha00}
 A.~Chakraborti, and B.K.~Chakrabarti, {Statistical
  mechanics of money: how saving propensity affects its distribution}, \emph{Eur. Phys. J. B},
  \textbf{17}: 167--170, 2000.


\bibitem{ChChSt05} A.~Chatterjee, B.K.~Chakrabarti, and
  R.B.~Stinchcombe,  {Master equation for a kinetic model of trading
  market and its analytic solution}, \emph{Phys. Rev. E}, \textbf{72}: 026126, 2005.

\bibitem{CPP}
S. Cordier, L. Pareschi and C. Piatecki, Mesoscopic modelling of financial markets, \emph{J. Stat. Phys.} \textbf{134}:161--184, 2009.  

\bibitem{CPT05} S. Cordier, L. Pareschi, and G. Toscani. On a kinetic model for a simple market economy. \emph{J. Stat. Phys.} \textbf{120}: 253--277, 2005.

\bibitem{DY} A. Dr\u{a}gulescu, and V.M. Yakovenko, {Statistical mechanics of money}, \emph{Eur. Phys. Jour. B} \textbf{17}: 723--729, 2000.

\bibitem{DMT}
B.~D{\"u}ring, D.~Matthes, and  G.~Toscani,  {Kinetic equations modelling wealth redistribution: a comparison of approaches}, \emph{Phys. Rev. E},
\textbf{78}:  056103, 2008.

\bibitem{DMT1}
B.~D{\"u}ring, D.~Matthes, and G.~Toscani, {A Boltzmann-type approach to the formation of wealth distribution curves},  (Notes of the Porto Ercole
School, June 2008),   \emph{Riv. Mat. Univ. Parma}, \textbf{8}: 199--261, 2009.

\bibitem{DS} 
B. D\"uring, N. Georgiou, and E. Scalas, {A stylised model for wealth distribution}, in \emph{Economic Foundations for Social Complexity Science}, Y. Aruka and
A. Kirman eds. 135--157, Springer, Singapore 2017.

\bibitem{DPT}
B. D\"uring, L. Pareschi, and  G. Toscani,  {
Kinetic models for optimal control of wealth inequalities}.  
\emph{Eur. Phys. J. B} \textbf{91}, Paper No. 265, 12 pp. 91B15, 2018.

\bibitem{DT2}
B. D\"uring, and  G. Toscani,  {International and domestic trading  and wealth distribution}, \emph{Commun. Math. Sci.}, \textbf{6}:1043--1058, 2008.

\bibitem{macro} M.S. Eichenbaumz, S. Rebelox, M. Trabandt, {The macroeconomics of epidemics}, \emph{NBER Working Paper} No. 26882, 2020.

\bibitem{FPTT16}
G.~Furioli, A.~Pulvirenti, E.~Terraneo, and G.~Toscani,
{Fokker--Planck equations in the modelling of socio-economic phenomena}, \emph{Math. Mod. Meth. Appl. Scie.}, \textbf{27}: 115--158, 2017.

\bibitem{GKLO} M. Gallegati, S. Keen, T. Lux, P. Ormerod,
Worrying trends in econophysics, \emph{
Physica A: Statistical Mechanics and its Applications},
 \textbf{370}:1--6, 2006.

\bibitem{GSV}
U. Garibaldi, E. Scalas, and P. Viarengo,
{Statistical equilibrium in simple exchange games II. The redistribution game},
\emph{Europ. Phys. J. B} \textbf{60}: 241--246, 2007.

\bibitem{GLN14}
A. Goenka a, L. Liu,  M-H. Nguyen, Infectious diseases and economic growth, \emph{Journal of Mathematical Economics}  \textbf{50}:34--53, 2014.

\bibitem{GCC16}
A. Ghosh, A. Chatterjee, J-I Inoue, B.K. Chakrabarti, Inequality measures in kinetic exchange models of wealth distributions, \emph{Physica A}, \textbf{451}:465--474,  2016.

\bibitem{GT-ec}
S. Gualandi and G. Toscani, Pareto tails in socio-economic phenomena: a kinetic description. \emph{Economics} \textbf{12}:1--17, 2018.

\bibitem{Gup}
A.K. Gupta, Models of wealth distributions: a perspective. In \emph{Econophysics and sociophysics: trends and perspectives}
 B.K. Chakrabarti, A. Chakraborti, A. Chatterjee (Eds.) Wiley VHC, Weinheim,  161--190, 2006.


\bibitem{HWH00} H.W. Hethcote. The mathematics of infectious diseases. {\em SIAM Review} \textbf{42}:599--653, 2000.

\bibitem{Kat}
G. Katriel, 
{Directed random market: the equilibrium distribution}, \emph{Acta Appl. Math.},
\textbf{139}: 95--103, 2015.

\bibitem{KM05} A. Korobeinikov,  P. K. Maini. Non-linear incidence and stability of infectious disease models. {\em Math. Med. and Bio.: A Journal of the IMA} \textbf{22}: 113--128, 2005. 

\bibitem{MT08}
D. Matthes and G. Toscani, On steady distributions of kinetic models of conservative economies. \emph{J. Stat.
Phys.}  \textbf{130}:1087--1117, 2008-.

\bibitem{PR}
L. Pareschi, and G. Russo. An introduction to Monte Carlo method for the Boltzmann equation. \emph{ESAIM: Proc.}, 10:35--75, 2001. 

\bibitem{PT-ss}
L. Pareschi, and G. Toscani,  {Self-similarity and power-like tails in nonconservative kinetic models}.
 \emph{J. Stat. Phys.}  \textbf{124}:747--779, 2006.

\bibitem{PT13}
L.~Pareschi, and G.~Toscani,
\newblock \emph{Interacting multiagent systems. Kinetic equations \& Monte Carlo
  methods},
\newblock (Oxford University Press, Oxford, 2013).

\bibitem{PT-k}
L. Pareschi, and G. Toscani,  {Wealth distribution and collective knowledge. A Boltzmann approach}, \emph{Phil. Trans. R. Soc. A}, \textbf{372}:20130396, 2014.

\bibitem{Par}
V. Pareto,
  \emph{Cours d'{\'E}conomie Politique},  Lausanne and Paris (1897).
  

\bibitem{SGD}
E. Scalas,U. Garibaldi, and S. Donadio, {Statistical equilibrium in simple exchange games I}
\emph{Europ. Phys. J. B}, \textbf{53}: 267--272, 2006.
  
  
\bibitem{Sl04}
F.~Slanina, {Inelastically scattering particles and wealth distribution in an open eco\-no\-my}, \emph{Phys. Rev. E} \textbf{69}:  046102, 2004.  

\bibitem{To1}
G. Toscani, Kinetic models of opinion formation, \emph{Commun.
Math. Sci.} \textbf{4}: 481--496, 2006.

\bibitem{TBD} 
G. Toscani, C. Brugna and S. Demichelis, Kinetic models for the trading of goods, \emph{J. Stat. Phys}, \textbf{151}: 549--566, 2013.


\bibitem{TTZ} 
G. Toscani, A. Tosin, M. Zanella. Multiple-interaction kinetic modelling of a virtual-item gambling economy. \emph{Phys. Rev. E}, 100(1): 012308, 2019. 

\bibitem{ZHJ} D. Zhang, M. Hu and Q. Ji, Financial markets under the global pandemic of COVID-19. \emph{Finance Research Letters}, 101528, 2020. Advance online publication. https://doi.org/10.1016/j.frl.2020.101528.

%%%%%%%%%%%%%%%%%%%%%%%%%%%%%%%%%


\end{thebibliography}
\end{document}